\documentclass[letterpaper,11pt]{article}
\pdfoutput=1
\pdfoutput=1

\usepackage{jheppub}
\usepackage{graphicx}
\usepackage{amsmath}
\usepackage{subfig}
\usepackage{xspace}
\usepackage{slashed}
\usepackage{xcolor}

\usepackage{bbm}

\newcommand{\df}{\mathrm{d}}

\usepackage{braket}
\usepackage{amssymb}
\usepackage{amsmath}
\usepackage{mathtools}

\def\nn{{\nonumber}}

\newcommand{\Tau}{\mathcal{T}}

\newcommand{\Ord}[1]{\mathcal{O}\left(#1\right)}

\def\cB{\mathcal{B}}

\def\cJ{\mathcal{J}}

\def\cM{\mathcal{M}}

\def\cO{\mathcal{O}}

\def\cP{\mathcal{P}}
\def\cJ{\mathcal{J}}
\def\cW{\mathcal{W}}

\def\cY{\mathcal{Y}}

\def\e{\epsilon}

\newcommand{\dbar}{d\hspace*{-0.08em}\bar{}\hspace*{0.1em}}

\newcommand{\cusp}{\mathrm{cusp}}

\newcommand{\kin}{\mathrm{kin}}
\newcommand{\hard}{\mathrm{hard}}

\newcommand{\mdm}{\mu\frac{d}{d \mu}}
\newcommand{\dmmp}{\frac{\df \mu'}{\mu'}}

\newcommand{\daap}{\frac{\df \alpha_s'}{(\alpha_s')^2}}

\newcommand{\Erf}{\mathrm{Erf}}
\newcommand{\Erfi}{\mathrm{Erfi}}

\newcommand{\LL}{\text{LL}}

\newcommand{\columnspinor}[2]{\colvec{2}{#1}{#2}}

\renewcommand{\smallmatrix}[4]{\begin{pmatrix}
    #1 & #2 \\
	#3 & #4 
\end{pmatrix}}
\newcount\colveccount
\newcommand*\colvec[1]{
        \global\colveccount#1
        \begin{pmatrix}
        \colvecnext
}
\def\colvecnext#1{
        #1
        \global\advance\colveccount-1
        \ifnum\colveccount>0
                \\
                \expandafter\colvecnext
        \else
                \end{pmatrix}
        \fi
}

\def\tr{{\rm tr}}

\newcommand\bn{{\bar n}}

\def\be{\begin{equation}}
\def\ee{\end{equation}}

\newcommand{\fd}[2]{\parbox{#1}{\includegraphics[width=#1]{#2}}}

\DeclareRobustCommand{\Sec}[1]{Sec.~\ref{#1}}

\DeclareRobustCommand{\App}[1]{App.~\ref{#1}}
\DeclareRobustCommand{\Tab}[1]{Table~\ref{#1}}

\DeclareRobustCommand{\Fig}[1]{Fig.~\ref{#1}}

\DeclareRobustCommand{\Eq}[1]{Eq.~(\ref{#1})}
\DeclareRobustCommand{\Eqs}[2]{Eqs.~(\ref{#1}) and (\ref{#2})}

\begin{document}

\preprint{MIT-CTP 4978}

\title{First Subleading Power Resummation for Event Shapes}

\author[1,2]{Ian Moult,}
\emailAdd{ianmoult@lbl.gov}
\author[3]{Iain W. Stewart,}
\emailAdd{iains@mit.edu}
\author[3]{Gherardo Vita,}
\emailAdd{vita@mit.edu}
\author[4]{and Hua Xing Zhu}
\emailAdd{zhuhx@zju.edu.cn}

\affiliation[1]{Berkeley Center for Theoretical Physics, University of California, Berkeley, CA 94720, USA}
\affiliation[2]{Theoretical Physics Group, Lawrence Berkeley National Laboratory, Berkeley, CA 94720, USA}
\affiliation[3]{Center for Theoretical Physics, Massachusetts Institute of Technology, Cambridge, MA 02139, USA}
\affiliation[4]{Department of Physics, Zhejiang University, Hangzhou, Zhejiang 310027, China\vspace{0.5ex}}

\abstract{We derive and analytically solve renormalization group (RG) equations of gauge invariant non-local Wilson line operators which resum logarithms for event shape observables $\tau$ at subleading power in the $\tau\ll 1$ expansion. These equations involve a class of universal jet and soft functions arising through operator mixing, which we call $\theta$-jet and $\theta$-soft functions. An illustrative example involving these operators is introduced which captures the generic features of subleading power resummation, allowing us to derive the structure of the RG to all orders in $\alpha_s$, and provide field theory definitions of all ingredients. As a simple application, we use this to obtain an analytic leading logarithmic result for the subleading power resummed thrust spectrum for  $H\to gg$ in pure glue QCD. This resummation determines the nature of the double logarithmic series at subleading power, which we find is still governed by the cusp anomalous dimension. We check our result by performing an analytic calculation up to ${\cal O}(\alpha_s^3)$. Consistency of the subleading power RG relates subleading power anomalous dimensions, constrains the form of the $\theta$-soft and $\theta$-jet functions, and implies an exponentiation of higher order loop corrections in the subleading power collinear limit. Our results provide a path for carrying out systematic resummation at subleading power for collider observables.}

\maketitle

\section{Introduction}\label{sec:intro}

Due to the complexity of interacting gauge theories in four dimensions, simplifying limits such as the soft, collinear, or Regge limits play a central role.  These limits are important both phenomenologically, where they often capture dominant contributions to processes of interest, as well as theoretically, where they place important constraints on the structure of amplitudes and cross sections. While well understood at leading power, less is known about the all orders perturbative structure of the subleading power corrections to these limits. These subleading power corrections have recently been attracting a growing level of interest, see for example \cite{Manohar:2002fd,Beneke:2002ph,Pirjol:2002km,Beneke:2002ni,Bauer:2003mga,Hill:2004if,Lee:2004ja,Dokshitzer:2005bf,Trott:2005vw,Laenen:2008ux,Laenen:2008gt,Paz:2009ut,Benzke:2010js,Laenen:2010uz,Freedman:2013vya,Freedman:2014uta,Bonocore:2014wua,Larkoski:2014bxa,Bonocore:2015esa,Bonocore:2016awd,Moult:2016fqy,Boughezal:2016zws,DelDuca:2017twk,Balitsky:2017flc,Moult:2017jsg,Goerke:2017lei,Balitsky:2017gis,Beneke:2017ztn,Feige:2017zci,Moult:2017rpl,Chang:2017atu}. A subset of these analyses consider power corrections to the threshold limit of Drell Yan and related processes, where there are no contributions from power corrections due to real collinear radiation. 

In this paper we will study the all orders structure of subleading power corrections to both the soft and collinear limits. This requires corrections beyond the type that can be studied from the threshold limit.  Using soft collinear effective theory (SCET) \cite{Bauer:2000ew, Bauer:2000yr, Bauer:2001ct, Bauer:2001yt}, which allows for a systematic power expansion using operator and Lagrangian based techniques, we will show for the first time how subleading power logarithms can be resummed to all orders in $\alpha_s$ for an event shape, which for concreteness we take to be thrust, $T=1-\tau$ \cite{Farhi:1977sg}, with $\tau\ll1$ in the simplified example of pure glue QCD for the process $H\to gg$ mediated by the effective operator $H G_{\mu\nu}^a G^{\mu\nu a}$ obtained by integrating out the top quark. In particular, we will show that at subleading power higher order corrections in $\alpha_s$ exponentiate at leading logarithmic (LL) accuracy into a single logarithmic term multiplying the same type of Sudakov form factor \cite{Sudakov:1954sw} as at leading power. 
Our approach is general, allowing other observables to be considered, and making clear what ingredients are needed to achieve higher logarithmic accuracy, as well as higher orders in the power expansion.

The all orders cross section for the thrust observable can be expanded in powers of $\tau$ (here $\tau$ is taken to be dimensionless), keeping all orders in $\alpha_s$ at each power
\begin{align}\label{eq:intro_expansion}
\frac{\df\sigma}{\df\tau} &=\frac{\df\sigma^{(0)}}{\df\tau} +\frac{\df\sigma^{(1)}}{\df\tau} +\frac{\df\sigma^{(2)}}{\df\tau}+\frac{\df\sigma^{(3)}}{\df\tau} +{\cal O}(\tau)\,.
\end{align}
Here $\df\sigma^{(n)}/\df\tau$ captures to all orders in $\alpha_s$ terms that scale like $\tau^{n/2-1}$, and for thrust the odd powers $d\sigma^{(2\ell+1)}/d\tau$ vanish. The leading power (LP) terms scale as $1/\tau$ (including $\delta(\tau)$) modulo logarithms. Explicitly, we have
\begin{align}
\frac{1}{\sigma_0}\frac{d\sigma^{(0)}}{d\tau} = \sum\limits_{n=0}^\infty \sum\limits_{m=-1}^{2n-1} \left(  \frac{\alpha_s(\mu)}{4\pi} \right)^n  c^{(0)}_{n,m}  {\cal L}_m(\tau)\,,
\end{align}
where ${\cal L}_{m\ge 0}(\tau) = [\theta(\tau)\log^m(\tau)/\tau]_+$ is a standard plus-function which integrates to zero over the interval $\tau\in [0,1]$, and ${\cal L}_{-1}(\tau) = \delta(\tau)$. Here the $c^{(0)}_{n,m}$ coefficients include $\log(\mu/Q)$ dependence, where $Q=m_H$ is the mass of the Higgs boson setting the scale of the hard scattering.
 All orders factorization theorems~\cite{Collins:1981uk,Collins:1985ue,Collins:1988ig,Collins:1989gx} can be proven at leading power for a number of event shape like observables~\cite{Sterman:1986aj,Bauer:2001yt,Bauer:2002nz,Fleming:2007qr,Schwartz:2007ib}. For the particular case of thrust in $H\to gg$, we have~\cite{Korchemsky:1999kt,Fleming:2007qr,Schwartz:2007ib}
\begin{align} \label{eq:fact0}
  \frac{1}{\sigma_0} \frac{\df\sigma^{(0)}}{\df\tau} 
   &=H^{(0)}(Q,\mu)\, \!\!\int\!\! ds_n ds_\bn dk \, \hat \delta_\tau\, 
     J^{(0)}_{g}(s_n,\mu ) ~ J^{(0)}_{g}(s_\bn,\mu) ~  S_{g}^{(0)}(k,\mu)  \,,
\end{align}
where 
\begin{align}
\hat \delta_\tau =\delta\left(\tau- \frac{s_n}{Q^2} -\frac{s_\bn}{Q^2} -\frac{k}{Q}\right)\,,
\end{align}
is the thrust measurement function.
Here $H^{(0)}(Q,\mu)$ is a hard function, $J^{(0)}_{g}(s,\mu )$ are gluon jet functions, and $S_{g}^{(0)}(k,\mu)$ is the adjoint soft function, whose precise definitions will be given in \Eqs{eq:jet_func}{eq:s_func} respectively. We normalize such that at lowest order $H^{(0)}$ is $1$, and the jet and soft functions are $\delta$-functions. The jet and soft functions are gauge invariant infrared finite matrix elements, which obey  simple renormalization group (RG) evolution equations that predict infinite towers of higher order logarithmically enhanced terms. The number of logarithms that are predicted is dictated by the logarithmic accuracy, denoted by N$^k$LL. Explicitly, for the first few orders, a resummation at N$^k$LL can be used to predict all the terms $c^{(0)}_{n,m}$, satisfying
\begin{align} \label{eq:whichms}
\text{LL predicts}: m=2n-1\,, \\
\text{NLL predicts}: m\geq 2n-2\,, \nn \\
\text{NNLL predicts}: m\geq 2n-4\,, \nn \\
\text{N$^3$LL predicts}: m\geq 2n-6\,, \nn 
\end{align}
for any $n$. Technically, for these resummations this counting is applied for $\log(d\sigma^{(0)}/dy)$ where $y$ is Fourier conjugate to $\tau$.\footnote{The standard counting which defines the resummation orders in position space is given by identifying the terms as $\log(d\sigma^{(0)}/dy)\simeq \sum_k  (\alpha_s \log)^k \log |_{\rm LL} + (\alpha_s\log)^k|_{\rm NLL} +  (\alpha_s\log)^k \alpha_s|_{\rm NNLL}+(\alpha_s\log)^k\alpha_s^2|_{\rm N^3LL}+\ldots$. This means that the resummation yields terms beyond those indicated in \Eq{eq:whichms} when expanded at the cross section level.} For thrust, these logarithms were first resummed to NLL in \cite{Catani:1991kz,Catani:1992ua}. Factorization and renormalization has been used to resum large logarithmic contributions to a number of $e^+e^-$ event shapes at leading power at N$^3$LL order \cite{Becher:2008cf,Abbate:2010xh,Chien:2010kc,Hoang:2014wka,Moult:2018jzp}.

Additional terms in \Eq{eq:intro_expansion} are suppressed by powers of $\lambda \sim \sqrt{\tau}$, with odd powers, $d\sigma^{(2\ell+1)}/d\tau$ vanishing, so that the series involves only integer powers of $\tau$ \cite{Beneke:2003pa,Lee:2004ja,Freedman:2013vya,Feige:2017zci}. These power suppressed terms do not involve distributions, and at power $\tau^{\ell-1}$ for $\ell \geq 1$ can be written as
\begin{align}\label{eq:subl_expansion}
\frac{1}{\sigma_0}\frac{d\sigma^{(2\ell)}}{d\tau} 
  =\sum\limits_{n=1}^\infty \sum\limits_{m=0}^{2n-1} \left(  \frac{\alpha_s(\mu)}{4\pi} \right)^n c^{(2\ell)}_{n,m}\, \tau^{\ell-1}\,\log^{m}(\tau)\,.
\end{align}
The structure of the subleading power terms is much less well understood, despite considerable effort. The first non-trivial power corrections are  described by $\df\sigma^{(2)}/\df\tau$, i.e. at  $\cO(\lambda^2)\sim \cO(\tau)$, which we will refer to as next-to-leading power (NLP). The subleading power terms at $\cO(\lambda^2)$ have recently been analytically computed in fixed order to $\cO(\alpha_s^2 \log^3)$ for thrust \cite{Freedman:2013vya,Moult:2016fqy,Boughezal:2016zws} and $N$-jettiness \cite{Moult:2016fqy,Boughezal:2016zws,Moult:2017jsg} for the first time, and the next-to-leading logarithms for $N$-jettiness at $\cO(\alpha_s)$ have been examined in \cite{Boughezal:2018mvf}. There has also been recent work on calculations of power corrections for $p_T$ in Drell-Yan \cite{Balitsky:2017flc,Balitsky:2017gis}, in the Regge limit \cite{Moult:2017xpp,Bruser:2018jnc}, and for subleading power quark mass effects \cite{Liu:2017vkm}. All these calculations have hinted at a simple structure for the power corrections, motivating an all orders understanding.

In a series of papers, we have developed within SCET all the ingredients relevant for the factorization and all orders description at $\cO(\lambda^2)$ for the case of dijet production from a color singlet current. This includes the bases of hard scattering operators \cite{Feige:2017zci,Moult:2017rpl,Chang:2017atu}, the factorization of the measurement function \cite{Feige:2017zci}, and the factorization of `radiative' contributions arising from subleading power Lagrangian insertions \cite{Moult:2019mog}. In this paper we combine these ingredients, and carry out the resummation of the leading logarithmic (LL) contributions to all orders in $\alpha_s$ for NLP corrections to thrust. In particular, this determines all terms $c^{(2)}_{n,2n-1}$ for any $n$ in \Eq{eq:subl_expansion}, giving all the terms in the series
\begin{align}
\frac{1}{\sigma_0}\frac{d\sigma^{(2)}}{d\tau}&=\left(  \frac{\alpha_s}{4\pi} \right)\, c_{1,1}^{(2)} \log \tau +\left(  \frac{\alpha_s}{4\pi} \right)^2\, c_{2,3}^{(2)} \log^3 \tau+ \left(  \frac{\alpha_s}{4\pi} \right)^3\, c_{3,5}^{(2)} \log^5 \tau +\cdots \,, \\
&=\left( \frac{\alpha_s}{4\pi} \right) 8 C_A \log\tau -
  \left(\frac{\alpha_s}{4\pi}\right)^2
32 C_A^2 \log^3 \tau + \left(\frac{\alpha_s}{4\pi}\right)^3 64 C_A^3
  \log^5 \tau + \ldots \,, \nn
\end{align}
where in the second line we have given the first few terms of the result that we will derive for thrust in pure glue $H\to gg$.
Note that this series starts at $\alpha_s \log \tau$, which has interesting consequences for the resummation. We will show that this necessitates the introduction of new jet and soft functions which arise through mixing, and which we term $\theta$-jet and $\theta$-soft functions. We will analytically solve the corresponding subleading power RG equation involving the mixing, and including the running coupling. We consider for simplicity the case of thrust in $H\to gg$ without fermions, i.e. in a pure SU$(3)$ Yang-Mills theory without matter. This will allow us to illustrate the conceptual complexities of renormalization at the cross section level in the simplest possible setting with a smaller set of operators. The addition of operators relevant for including fermions will be considered in future work.

An outline of this paper is as follows. In \Sec{sec:renorm} we show in the context of an illustrative example how one can renormalize subleading power jet and soft functions. The illustrative example allows for an understanding of the renormalization to all orders in $\alpha_s$, and allows us to provide complete field theoretical definitions for all functions involved in the RG flow. This involves a new class of jet and soft functions which arise at cross section level through mixing, which we demonstrate is a generic feature at subleading power that is needed to predict the series that starts at $\alpha_s \log \tau$. At $\cO(\lambda^2)$, this gives rise to a $2\times 2$ mixing structure for the RG equations. We study in detail the consistency equations for this type of RG evolution, allowing us to derive powerful and general constraints on the structure of operators that can be mixed into at subleading powers. In \Sec{sec:solution} we solve the general form of the subleading power mixing equation, including the running coupling as is relevant for subleading power resummation in QCD. In \Sec{sec:fact} we apply this to resum the leading logs at subleading power for thrust in pure glue $H\to gg$, deriving the structure of the Sudakov exponent for the subleading power corrections. In \Sec{sec:split} we perform a fixed order check of our result. We explicitly calculate to $\cO(\alpha_s^3)$ the $\cO(\lambda^2)$ leading logarithms, confirming the result predicted by the RG. Furthemore, we interpret the fixed order expansion in terms of information about the $\cO(\alpha_s^n)$ corrections to subleading power splitting functions.  We conclude in \Sec{sec:conc}.

\section{Renormalization at Subleading Power}\label{sec:renorm}

In this section we study the structure and completeness of jet and soft functions for renormalization group equations at subleading power.   In \Sec{sec:illexample} we introduce a simple illustrative example which can be studied to all orders from known factorization properties at leading power, and from which many interesting lessons about the structure of subleading power resummation can be deduced.  This example also appears explicitly for thrust in $H\to gg$ from contributions from subleading power kinematic corrections.  In \Sec{sec:renorm_a}, we show that the renormalization of the subleading power jet and soft functions in our illustrative example leads to mixing into jet and soft functions involving $\theta$-functions of the measurement operator, which we term $\theta$-jet and $\theta$-soft functions, and we derive the structure of the RG to all orders in $\alpha_s$.  In \Sec{sec:consistency} we study RG consistency in a setup that is a generalization of our illustrative example in order to derive general constraints at subleading power on the structure of anomalous dimensions and on the appearance of $\theta$-function operators.

\subsection{An Illustrative Example at Subleading Power}\label{sec:illexample}

Our illustrative example of a subleading power factorization is obtained by multiplying the leading power factorization by $\tau$ and using
\begin{align}
  \tau \hat \delta_\tau = \tau \delta(\tau - \tau_n -\tau_\bn - \tau_s) =  (\tau_n+\tau_\bn+\tau_s)    \hat \delta_\tau \,,
\end{align}
which gives a subleading power cross section whose factorized structure follows immediately from the leading power factorization of \Eq{eq:fact0}:
\begin{align} \label{eq:fact_NLP_multTau}
 \frac{1}{\sigma_0} \frac{\df\sigma^{(2)}}{\df\tau} 
   &= H^{(0)}(Q,\mu) \int \frac{ds_n ds_\bn dk}{Q^2}\, \hat \delta_{\tau}\, \left[ s_n  J^{(0)}_{g}(s_n,\mu ) \right]  J^{(0)}_{g}(s_\bn,\mu) S_{g}^{(0)}(k,\mu)    \\
   &+ H^{(0)}(Q,\mu) \int \frac{ds_n ds_\bn dk}{Q^2} \,\hat \delta_{\tau} \, J^{(0)}_{g}(s_n,\mu )  \left[ s_\bn  J^{(0)}_{g}(s_\bn,\mu) \right ]   S_{g}^{(0)}(k,\mu)  \nn \\
   &+ H^{(0)}(Q,\mu)  \int \frac{ds_n ds_\bn dk}{Q} \,  \hat \delta_{\tau}\, J^{(0)}_{g}(s_n,\mu )  J^{(0)}_{g}(s_\bn,\mu)    \left[ k  S_{g}^{(0)}(k,\mu) \right]   \,.\nn
\end{align}
This can be written in terms of subleading power jet and soft functions as
\begin{align}  \label{eq:fact_NLP_multTau_rewrite}
  \frac{1}{\sigma_0} \frac{\df\sigma^{(2)}}{\df\tau}   
      &= H^{(0)}(Q,\mu)\int \frac{ds_n ds_\bn dk}{Q^2}\, \hat \delta_\tau\,  J^{(2)}_{g}(s_n,\mu )  J^{(0)}_{g}(s_\bn,\mu) S_{g}^{(0)}(k,\mu)    \\
   &+ H^{(0)}(Q,\mu) \int\frac{ds_n ds_\bn dk}{Q^2}\, \hat \delta_\tau\,  J^{(0)}_{g}(s_n,\mu )   J^{(2)}_{g}(s_\bn,\mu)   S_{g}^{(0)}(k,\mu)  \nn \\
   &+ H^{(0)}(Q,\mu) \int \frac{ds_n ds_\bn dk}{Q} \,  \hat \delta_\tau\, J^{(0)}_{g}(s_n,\mu )  J^{(0)}_{g}(s_\bn,\mu)     S_{g}^{(2)}(k,\mu)\,. \nn
\end{align}
The superscripts indicate the power of the function, namely those with superscript $(0)$ are LP in the $\tau$ expansion, while  those with superscript $(2)$ are power suppressed by $\lambda^2 \sim \tau$.
In this factorization, $H^{(0)}(Q,\mu)$ is the leading power hard function, which is process dependent, and will not play an important role in the current discussion.
The leading power jet function, which for $H\to gg$ is a gluon jet function, is defined as a matrix element of collinear fields
\begin{align}\label{eq:jet_func}
J^{(0)}_{g}(s,\mu)
  &= \frac{(2\pi)^3}{(N_c^2-1)}\Big\langle 0 \Big|\, \cB^{a}_{n\perp\mu} (0)\,\delta(Q+\bar \cP) \delta^2(\cP_\perp)\, \delta\left(\frac{s}{Q}-\hat \Tau\right)\, \cB^{\mu a}_{n\perp}(0) \,\Big|0\Big\rangle
\,, 
\end{align}
where $\cB^{a\mu}_{n\perp}$, is a gauge invariant gluon field (see \Eq{eq:softgluondef} for an explicit definition), and the leading power adjoint soft function is given by
\begin{align}\label{eq:s_func}
S_{g}^{(0)}(k,\mu)=\frac{1}{(N_c^2-1)}  \tr \big\langle 0 \big| \cY^T_{\bar n}(0) \cY_n(0) \delta(k-\hat \Tau)  \cY_n^T(0) \cY_{\bar n}(0) \big|0 \big\rangle\,,
\end{align}
where $\cY_n$, $\cY_\bn$  are adjoint Wilson lines along the given lightlike directions. Explicitly,
\be\label{eq:Wilson_def}
\cY^{bc}_n(x)=\mathbf{P} \exp \left [ g \int\limits^{\infty}_0 ds\, n\cdot A^a_{us}(x+sn) f^{abc}\right]\,.
\ee
In both cases, $\hat \Tau$ is an operator that returns the value of $\Tau$ measured on a given state, where the dimensionless thrust $\tau=\Tau/Q$. In general it can be written in terms of the energy momentum tensor of the effective theory \cite{Lee:2006nr,Sveshnikov:1995vi,Korchemsky:1997sy,Bauer:2008dt,Belitsky:2001ij,Mateu:2012nk}. At tree level, $J^{(0)}_{g}(s,\mu)=\delta(s)+\cO(\alpha_s)$ and $S_{g}^{(0)}(k,\mu)=\delta(k)+\cO(\alpha_s)$.

After multiplying by $\tau$, the operator definitions for the subleading power jet and soft functions appearing in \Eq{eq:fact_NLP_multTau} are simply
\begin{align}\label{eq:tau_funcs}
J^{(2)}_{g,  \delta}(s,\mu)&=\frac{(2\pi)^3}{(N_c^2-1)}   \Big\langle 0 \Big|\, \cB^{\mu a}_{n\perp} (0)\,\delta(Q+\bar \cP) \delta^2(\cP_\perp)\, s\,\delta\left(\frac{s}{Q}-\hat \Tau\right)\, \cB^{\mu a}_{n\perp,\omega}(0) \,\Big|0\Big\rangle\,, \\
S^{(2)}_{g, \delta}(k,\mu)&= \frac{1}{(N_c^2-1)}\tr \langle 0 |  \cY^T_{\bar n}(0) \cY_n(0)\, k ~\delta(k-\hat \Tau) \cY_n^T(0) \cY_{\bar n}(0) |0\rangle\,.\nn
\end{align}
 The subscript $\delta$ is meant to indicate that the measurement function that appears is the same as the leading power measurement. The mass dimension of both functions in \Eq{eq:tau_funcs} is zero. Although this example may appear too trivial, it turns out to become quite interesting when we consider the RG evolution of these subleading power jet and soft functions, which we do next.

\subsection{$\theta$-jet and $\theta$-soft Functions and RG Equations}\label{sec:renorm_a}

The RG for the subleading power jet and soft functions in \Eq{eq:tau_funcs} is easily deduced from the RG evolution of the leading power jet and soft functions. The leading power jet and soft functions satisfy the RG equations
\begin{align} \label{eq:RGES0J0}
\mu \frac{dS^{(0)}_{g}(k,\mu) }{d\mu} 
&=  \int\!\! dk' \: \gamma^S_g(k-k',\mu)\, S^{(0)}_{g}(k',\mu)\,,
\\
\mu \frac{dJ^{(0)}_{g}(s,\mu) }{d\mu}  &=  \int ds'  \gamma^J_{g}(s- s',\mu)\: J^{(0)}_{g}(s',\mu)\,, \nn
\end{align}
where the form of the anomalous dimensions to all orders in $\alpha_s$ is
\begin{align}\label{eq:LP_anom_dim}
 \gamma^S_g(k,\mu) &= 4 \Gamma_{\text{cusp}}^g[\alpha_s] \, \frac{1}{\mu} \biggl[  \frac{\mu\,\theta(k)}{k}  \biggr]_+  + \gamma_g^S[\alpha_s] \: \delta(k) \,,
  \\
  \gamma^J_g(s,\mu) &= -2 \Gamma_{\text{cusp}}^g[\alpha_s] \, \frac{1}{\mu^2} \biggl[  \frac{\mu^2\,\theta(s)}{s}  \biggr]_+  + \gamma_g^J[\alpha_s] \: \delta(s) \,,
   \nn
\end{align}
with $\Gamma^{g}_{\cusp}[\alpha_s]$ the gluon cusp anomalous dimension \cite{Korchemsky:1987wg,Korchemskaya:1992je}.

We can now derive the all orders result for the RG evolution of the subleading power jet and soft functions. Multiplying the leading power soft function by $k$, we find for the soft function
\begin{align}
\mu \frac{d}{d\mu} & k S^{(0)}_{g}(k,\mu) 
=  \int\!\! dk' \: \big((k-k')+k'\big)\gamma^S_g(k-k',\mu)\, S^{(0)}_{g}(k',\mu)\,,  \\
&=\int\!\! dk' \: \big((k-k')+k'\big)  \left\{  4 \Gamma_{\text{cusp}}^g[\alpha_s] \, \frac{1}{\mu} \biggl[  \frac{\mu\,\theta(k-k')}{k-k'}  \biggr]_+  + \gamma_g^S[\alpha_s] \: \delta(k-k')  \right\} S^{(0)}_{g}(k',\mu) \nn \\
&=\int\!\! dk'  4 \Gamma_{\text{cusp}}^g[\alpha_s] \theta(k-k') S^{(0)}_{g}(k',\mu)  + \int\!\! dk' \: \gamma^S_g(k-k',\mu)\, k' S^{(0)}_{g}(k',\mu)\,.\nn 
\end{align}
This implies
\begin{align}\label{eq:tau_mix}
\mu \frac{d}{d\mu}  S^{(2)}_{g,\delta}(k,\mu) &=4\Gamma^g_{\cusp}[\alpha_s] ~   S^{(2)}_{g,\theta}(k,\mu)    + \int\!\! dk' \: \gamma^S_g(k-k',\mu)\, S^{(2)}_{g,\delta}(k',\mu)\,.
\end{align}
Here we have defined the new power suppressed soft function
\begin{align}\label{eq:theta_soft_first}
S^{(2)}_{g,\theta}(k,\mu)&= \frac{1}{(N_c^2-1)} \tr \langle 0 | \cY^T_{\bar n} (0)\cY_n(0) \theta(k-\hat \Tau) \cY_n^T(0) \cY_{\bar n}(0) |0\rangle\,.
\end{align}
We refer to this as a $\theta$-soft function. Its tree level value is $S^{(2)}_{g,\theta}(k,\mu)=\theta(k) +\cO(\alpha_s)$.
This function receives its power suppression from its measurement function, $\theta(k-\hat \Tau)$. In particular, $\theta(\tau)\sim \cO(\tau^0)$, while $\delta(\tau) \sim \cO(1/\tau)$. 

Performing an identical exercise for the jet function, we obtain
\begin{align}\label{eq:tau_jet_mix}
\mu \frac{d}{d\mu}  J^{(2)}_{g,\delta}(s,\mu) 
&=-2\Gamma^g_{\cusp}[\alpha_s] ~   J^{(2)}_{g,\theta}(s,\mu)    + \int\!\! ds' \: \gamma^J_g(s-s',\mu)\, J^{(2)}_{g,\delta}(s',\mu)\,.
\end{align}
Here we have defined the subleading power jet function
\begin{align}\label{eq:theta_op}
J^{(2)}_{g,\theta}(s,\mu)&=\frac{(2\pi)^3}{(N_c^2-1)} \Big\langle 0 \Big|\, \cB^{\mu a}_{n\perp} (0)\,\delta(Q+\bar \cP) \delta^2(\cP_\perp)\, \theta\left(\frac{s}{Q}-\hat \Tau\right)\, \cB^{\mu a}_{n\perp,\omega}(0) \,\Big|0\Big\rangle\,,
\end{align}
which we will refer to as a $\theta$-jet function.  Its tree level value is $J^{(2)}_{g,\theta}(s,\mu)=\theta(s) +\cO(\alpha_s)$. In \cite{Paz:2009ut} it was also found that additional subleading power jet functions whose tree level values were $\theta$-functions were required due to the non-closure of the RG evolution, and it was conjectured that they took the form of \Eq{eq:theta_op}.   Our illustrative example has allowed us to derive the necessity of such operators in a straightforward manner, and prove that here this new function suffices to all orders in $\alpha_s$. More general constraints on the functions that can appear through mixing at subleading power will be derived from the consistency of the RG equations in \Sec{sec:consistency}.

Interestingly, we see that the evolution equation for the power suppressed jet and soft functions are no longer homogeneous evolution equations. In particular, they mix into the $\theta$-jet and $\theta$-soft functions. 
This clearly shows that a new class of subleading power operators, namely the $\theta$-jet and $\theta$-soft operators, are required to renormalize consistently at subleading power in SCET.
These operators do not appear  at amplitude level, but instead arise from mixing at cross section level. It is clear that they have all the correct symmetry properties, as well as the correct power counting, and therefore it is not unexpected that they can be generated by RG evolution.

The renormalization group evolution of the $\theta$-function operators can also be derived by integration of the leading power RG equation. Considering explicitly the soft function, we have
\begin{align}
\mu \frac{d}{d\mu}  S^{(2)}_{g,\theta}(k,\mu) &= \int dk' \theta(k-k') \int dk'' \gamma_g^{S}(k'-k'',\mu) S_g^{(0)}(k'',\mu)  \\
&= \int dk' \gamma_g^{S}(k-k',\mu) \int dk'' \theta(k'-k'')S_g^{(0)}(k'',\mu) \nn \\
&= \int dk' \gamma_g^{S}(k-k',\mu) S_{g,\theta}^{(2)}(k',\mu)\,.\nn
\end{align}
We therefore find that to all orders in $\alpha_s$, the RG for the $\theta$-jet and $\theta$-soft operators is identical to that of the leading power jet and soft functions
\begin{align}\label{eq:theta_RG}
\mu \frac{d}{d\mu}  S^{(2)}_{g,\theta}(k,\mu)& = \int dk' \gamma^S_g(k-k')S^{(2)}_{g,\theta}(k',\mu)\,, \\
\mu \frac{d}{d\mu}  J^{(2)}_{g,\theta}(s,\mu) &= \int ds' \gamma^J_{g}(s- s')J^{(2)}_{g,\theta}(s',\mu)\,.\nn
\end{align}
Together \Eqs{eq:tau_mix}{eq:tau_jet_mix} combined with \Eq{eq:theta_RG} give a simple, closed $2 \times 2$ matrix RG structure for the subleading power jet and soft functions
\begin{align}\label{eq:2by2mix}
\mu \frac{d}{d\mu} \Biggl(\begin{array}{c} J^{(2)}_{g, \delta}(s,\mu) \\[5pt]
  J^{(2)}_{g,\theta}(s,\mu) \end{array} \Biggr) &= \int ds' 
  \Biggl( \begin{array}{cc} 
\gamma^J_{g,\delta\delta}(s-s') \qquad &  \gamma^J_{g,\delta\theta}\, \delta(s-s') \\[5pt]
  0 &  \gamma^J_{g, \theta \theta}(s-s')    
\end{array} \Biggr) 
\Biggl(\begin{array}{c} J^{(2)}_{g,\delta}(s',\mu) \\[5pt] 
J^{(2)}_{g, \theta}(s',\mu) \end{array} \Biggr) \,, \\
\mu \frac{d}{d\mu}\left(\begin{array}{c} S^{(2)}_{g,\delta}(k,\mu) \\ S^{(2)}_{g,\theta}(k,\mu) \end{array} \right) &= \int dk' \left( \begin{array}{cc} \gamma^S_{g,\delta \delta}(k-k',\mu) & \gamma^S_{g,\delta \theta}\, \delta(k-k') \\  0 &  \gamma^S_{g, \theta \theta}(k-k',\mu)   \end{array} \right) \left(\begin{array}{c} S^{(2)}_{g,\tau\delta}(k',\mu) \\ S^{(2)}_{g, \theta}(k',\mu) \end{array} \right) \,. \nn
\end{align}
Fourier transforming to position space 
\begin{align} \label{eq:FT_JS}
\tilde J_x^{(2)}(y) &= \int\!\! ds~ e^{-is y}~  J_x^{(2)}(s) \,,
& \tilde S_x^{(2)}(z) &= \int\!\! dk~ e^{-ik z}~  S_x^{(2)}(k) \,,
\end{align}
(where here the mass dimensions are $[y]=-2$ and $[z]=-1$)
these RG equations become multiplicative
\begin{align}\label{eq:RGyz}
   \mdm \columnspinor{\tilde J^{(2)}_{g,\delta} (y,\mu)}{\tilde J^{(2)}_{g,\theta}(y,\mu)} 
   &= \smallmatrix{\tilde \gamma^J_{g,\delta\delta}(y,\mu)}{ \gamma^J_{g,\delta\theta}[\alpha_s]}{0}{\tilde \gamma^J_{g,\theta\theta}(y,\mu)} \columnspinor{\tilde J^{(2)}_{g,\delta}(y,\mu)}{\tilde J^{(2)}_{g,\theta}(y,\mu)} 
  \,, \\
   \mdm \columnspinor{\tilde S^{(2)}_{g,\delta} (z,\mu)}{\tilde S^{(2)}_{g,\theta}(z,\mu)} 
   &= \smallmatrix{\tilde \gamma^S_{g,\delta\delta}(z,\mu)}{ \gamma^S_{g,\delta\theta}[\alpha_s]}{0}{\tilde \gamma^S_{g,\theta\theta}(z,\mu)} \columnspinor{\tilde S^{(2)}_{g,\delta}(z,\mu)}{\tilde S^{(2)}_{g,\theta}(z,\mu)}
 \,. \nn
\end{align}

For our illustrative example, the RG equations in \Eq{eq:2by2mix} or \Eq{eq:RGyz} are  valid to all orders in $\alpha_s$, and we can identify that
\begin{align}\label{eq:anom_dim_mix_diag}
\gamma^S_{g,\delta \delta}(k, \mu)=  \gamma^S_{g, \theta\theta}(k,\mu)=\gamma_g^S(k,\mu)\,, \\
\gamma^J_{g,\delta \delta}(s, \mu)=  \gamma^J_{g, \theta \theta}(s,\mu)=\gamma_{g}^J(s,\mu) \,, \nn
\end{align}
where $\gamma^S_g(k,\mu)$ and $\gamma_{g}^J(s,\mu)$ are the LP anomalous dimensions in \Eq{eq:LP_anom_dim}. They include the cusp anomalous dimensions, and hence drive double logarithmic evolution. On the other hand, in our illustrative example the off diagonal terms in \Eq{eq:2by2mix} are
\begin{align}\label{eq:anom_dim_mix}
\gamma^S_{g,\delta \theta}&=  4\Gamma^g_{\cusp}[\alpha_s]\,,\\
\gamma^J_{g,\delta \theta}&= -2\Gamma^g_{\cusp}[\alpha_s]\,, \nn
\end{align}
which generate single logarithmic terms.

The particular relations for the anomalous dimensions of \Eqs{eq:anom_dim_mix_diag}{eq:anom_dim_mix}, and in particular the fact that the mixing anomalous dimension is proportional to the cusp anomalous dimension,  is a feature of this specific illustrative example, and will not in general be true. However, the general features of this example will be true at subleading power. In particular, subleading power jet and soft functions will exhibit single logarithmic mixing with $\theta$-function operators, and diagonal anomalous dimensions corresponding to operator self mixing will give rise to double logarithmic evolution.  In \Sec{sec:consistency} we will discuss more general constraints on the subleading power anomalous dimensions and the types of functions which can arise through mixing, from RG consistency constraints in SCET.

From this example we have shown how subleading power jet and soft functions involving $\theta$-function measurement operators arise in a straightforward manner,  we have derived their field structure to all orders in $\alpha_s$, and we have shown that their RG closes in a $2 \times 2$ form.  Before solving this subleading power RG equation, it is also useful to see how this mixing appears from the perspective of a fixed order calculation for the subleading power soft function. This will illustrate that this phenomenon of mixing is  generic at subleading power, due to the fact that subleading power corrections first contribute with a real emission without virtual corrections, and is not simply a feature of the specific example considered here.

At lowest order, the power suppressed soft function vanishes
\begin{align}
\left.S^{(2)}_{g,\delta} (k,\mu)\right|_{\cO(\alpha_s^0)}=\fd{2cm}{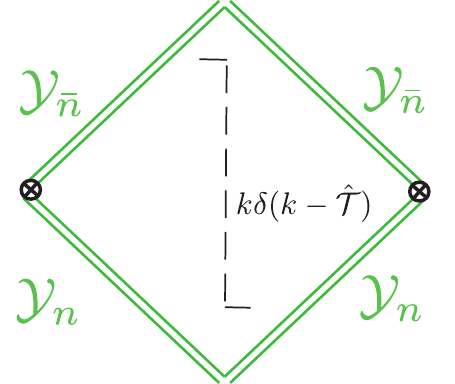}=k\delta(k)=0\,.
\end{align}
With a single emission, we have
\begin{align}
\left.S^{(2)}_{g,\delta} (k,\mu)\right|_{\cO(\alpha_s)}
  =2\fd{2cm}{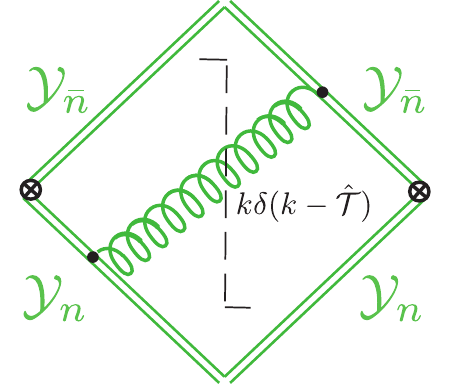}
  =4g^2  \left(  \frac{\mu^2 e^{\gamma_E}}{4\pi} \right)^\epsilon \! C_A\! \int \frac{d^d\ell}{(2\pi)^d} \frac{1}{\ell^+\ell^-} 2\pi\delta(\ell^2) \theta(\ell^0) k \delta(k-Q\hat \tau)\,,
\end{align}
where the measurement function on a single particle state is given by
\begin{align}
k \delta (k -Q\hat \tau)& =k \delta \left(  k -\ell^+\right) \theta(\ell^- -\ell^+) +k \delta \left(  k -\ell^-\right) \theta(\ell^+ -\ell^-)\nn \\
&=2k \delta \left( k -\ell^+ \right) \theta(\ell^- -\ell^+)\,,
\end{align}
using the $\ell^+\leftrightarrow \ell^-$ symmetry of this particular integrand.
Using the delta functions to perform the integrals of the $l_\perp$ and $l^+$, we find
\begin{align}\label{eq:NLP_soft_toy}
S^{(2)}_{g,\delta} (k,\mu)&=\frac{8\alpha_s C_A k^{-\epsilon}}{\Gamma(1-\epsilon)(4\pi)^{1-\epsilon}} \left(  \frac{\mu^2 e^{\gamma_E}}{4\pi} \right)^\epsilon   \int\limits_{k}^\infty \frac{d\ell^-}{2\pi} \frac{1}{(\ell^-)^{1+\epsilon}} =\frac{8\alpha_s C_A e^{\epsilon \gamma_E}}{\Gamma(1-\epsilon)(4\pi)} \left(  \frac{\mu^2 }{k^2} \right)^\epsilon \frac{1}{\epsilon}\nn \\
&=8C_A\frac{\alpha_s(\mu)}{4\pi}  \theta(k) \left(   \frac{1}{\epsilon} +  \log   \frac{\mu^2}{k^2} +\cO(\epsilon)\right)\,.
\end{align}
Here we clearly see that an SCET UV divergence from $\ell^-\to \infty$ appears at the first order at which this power suppressed soft function is non-vanishing.

Although we are considering a specific subleading power example, these two calculations illustrate a general phenomenon at subleading power: subleading power jet and soft functions vanish at lowest order since purely virtual corrections are leading power, scaling like $\delta(\tau)$, and they in general have a UV divergence in SCET at the first perturbative order at which they appear. Without the knowledge of the $\theta$-soft and $\theta$-jet operators, this behavior is confusing, since it is not clear what renormalizes this divergence. However, with an understanding of the presence of these $\theta$-function operators, we can now straightforwardly interpret the fixed order calculation of the subleading power soft function in \Eq{eq:NLP_soft_toy} as operator mixing, and immediately read off the anomalous dimension from the $1/\epsilon$ pole in the standard way.
The operator $S^{(2)}_{g,\theta}$ is non-zero at tree level, and simply gives
\begin{align}
\left. S^{(2)}_{g,\theta}(k,\mu)\right|_{\cO(\alpha_s^0)}=\fd{2cm}{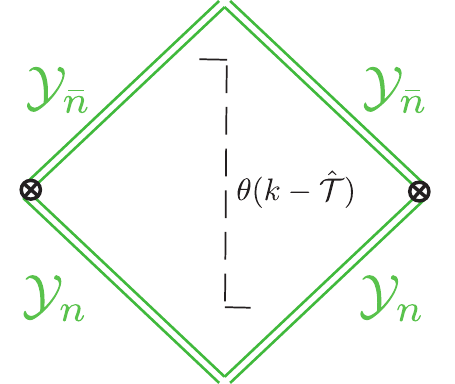} =\theta(k)\,.
\end{align} 
The renormalization of this operator provides the needed counterterm,
and from \Eq{eq:NLP_soft_toy} we find
\begin{align}
  \gamma^S_{g,\delta  \theta}&= 16 \frac{\alpha_s}{4\pi}C_A=  4\frac{\alpha_s}{4\pi}\Gamma^{g,0}_{\cusp}\,,
\end{align}
where $\Gamma^{g,0}_{\cusp}=4C_A$ is the one-loop gluon cusp anomalous dimension. This result is in agreement with our derivation from the known structure of the RG equations in \Eq{eq:tau_mix}. This example clearly resolves any confusion arising in the renormalization of the subleading power operators, which with the addition of subleading power $\theta$-jet and $\theta$-soft functions becomes a standard operator mixing problem.

\subsection{Renormalization Group Consistency}\label{sec:consistency}

Motivated by the structure of the RG equations in our illustrative example, we consider a somewhat more general factorization theorem where the soft and jet sectors have an analogous $2\times2$ mixing structure with some unknown functions that do not appear in the matching, without working under the assumption that these functions take the form of the $\theta$-jet or $\theta$-soft functions of the previous section. The fact that the cross section is $\mu$-independent implies RG consistency equations in SCET that yield relations between the anomalous dimensions of hard, jet, and soft functions, and will allow us to prove on general grounds that the functions appearing through mixing at subleading power must be integrals of the leading power functions in the factorization theorem. This shows that the $\theta$-jet or $\theta$-soft functions appear much more generally than in our illustrative example. It will also allow us to demonstrate that there will always be at least pairs of subleading power $\theta$-soft and $\theta$-collinear functions. 

We consider terms in a subleading power factorization theorem where the power corrections occur in either a jet or soft function with the form
\begin{align}\label{eq:eg_RG_consistency}
 \frac{1}{\sigma_0} \frac{\df\sigma^{(2)}}{\df\tau} 
   &= 2H_1(Q,\mu)\int\!  \frac{ds_n ds_\bn dk}{Q^2}\, \hat \delta_\tau\, J^{(2)}_{\delta}(s_n,\mu )  J^{(0)}(s_\bn,\mu) S^{(0)}(k,\mu)    \\
   &\ \
  + H_2(Q,\mu) \int\! \frac{ds_n ds_\bn dk}{Q} \,  \hat \delta_\tau\, J^{(0)}(s_n,\mu )  J^{(0)}(s_\bn,\mu)   S_{\delta}^{(2)}(k,\mu)  \,,\nn
\end{align}
where we have used the $n\leftrightarrow \bar n$ symmetry to write corrections to the two jet functions into a single expression. Here $H_{1}=1 +{\cal O}(\alpha_s)$ and $H_2=1+{\cal O}(\alpha_s)$ are taken to be dimensionless hard functions. We will assume that these $H_i$ do not mix, so $\mu\frac{d}{d\mu} H_i(Q,\mu)=\gamma_{H_i}(Q,\mu) H_i(Q,\mu)$.  We will also assume that $J^{(2)}_{\delta}$ and $S^{(2)}_{\delta}$ start at ${\cal O}(\alpha_s)$, and obey $2\times2$ mixing equations of the form in \Eq{eq:2by2mix} which has them mix with operators starting at ${\cal O}(\alpha_s^0)$. Importantly, here we do not assume that $J^{(2)}_{\delta}$ and $S^{(2)}_{\delta}$ are related to the functions defined in \Eq{eq:tau_funcs}. We also assume that the terms in \Eq{eq:eg_RG_consistency} close in the renormalization group flow (at least up to some order in the N$^k$LL expansion, though we will shortly focus on LL order). From \Eq{eq:fact_NLP_multTau_rewrite} we see that the expression for the cross section in our illustrative example satisfies all the above assumptions and is a special case of the assumed form. With the above assumptions, our goal is to derive RG consistency equations by demanding the RG invariance of this cross section, $\mu d/d\mu\, d\sigma^{(2)}/d\tau = 0$.

For the analysis of RG consistency it is most convenient to Fourier transform $\tau$ to position space, so that \Eq{eq:eg_RG_consistency} becomes
\begin{align}\label{eq:eg_RG_consistency_y}
 \frac{1}{\sigma_0} \frac{\df\sigma^{(2)}}{\df y}
   &\equiv \int\! d\tau \: e^{-i\tau y}   \frac{1}{\sigma_0} \frac{\df\sigma^{(2)}}{\df\tau}  \\
   &= \frac{2}{Q^2} H_1(Q,\mu) \tilde J^{(2)}_{\delta}\Bigl(\frac{y}{Q^2},\mu \Bigr) \tilde J^{(0)}\Bigl(\frac{y}{Q^2},\mu \Bigr)
   \tilde S^{(0)}\Bigl(\frac{y}{Q},\mu\Bigr)  
   \nn  \\
   &\ \
  + \frac{1}{Q} H_2(Q,\mu) \tilde J^{(0)}\Bigl(\frac{y}{Q^2},\mu \Bigr) \tilde J^{(0)}\Bigl(\frac{y}{Q^2},\mu\Bigr)  \tilde S_{\delta}^{(2)}\Bigl(\frac{y}{Q},\mu\Bigr) 
   \,.\nn
\end{align}
Here $y$ is dimensionless and the Fourier transforms of jet and soft functions are defined as in \Eq{eq:FT_JS}. Differentiating each of the terms in \Eq{eq:eg_RG_consistency_y} and using \Eq{eq:RGES0J0} and the analog of \Eq{eq:RGyz} gives terms involving anomalous dimensions times the same functions back again, plus the terms involving mixing into additional functions. For notational convenience we will refer to these as $\theta$-jet and $\theta$-soft functions, although we will not assume that they take the functional form of the illustrative example result in \Eqs{eq:theta_soft_first}{eq:theta_op}. We therefore arrive at the following consistency equation (here for brevity we suppress the $\mu$ arguments in all functions and anomalous dimensions), 
\begin{align}
0 &= \mu \frac{d}{d\mu} \left[ \frac{d\sigma^{(2)}}{\df y}  \right] 
  \\
 &= \frac{2}{Q^2} \biggl[ \gamma_{H_1}(Q) + \tilde\gamma_{\delta\delta}^J\Bigl(\frac{y}{Q^2}\Bigr) + 
 \tilde\gamma_{J^{(0)}}\Bigl(\frac{y}{Q^2}\Bigr)  +
 \tilde\gamma_{S^{(0)}}\Bigl(\frac{y}{Q}\Bigr) \biggr] 
  H_1(Q) \tilde J^{(2)}_{\delta}\Bigl(\frac{y}{Q^2}\Bigr) \tilde J^{(0)}\Bigl(\frac{y}{Q^2} \Bigr) \tilde S^{(0)}\Bigl(\frac{y}{Q}\Bigr)  
  \nn\\
 &\quad + \frac{2}{Q^2} \gamma_{\delta\theta}^J[\alpha_s] 
  H_1(Q) \tilde J^{(2)}_{\theta}\Bigl(\frac{y}{Q^2}\Bigr) \tilde J^{(0)}\Bigl(\frac{y}{Q^2} \Bigr) \tilde S^{(0)}\Bigl(\frac{y}{Q}\Bigr)  
  \nn\\ 
 & \quad + \frac{1}{Q} \biggl[ \gamma_{H_2}(Q) + 
  2 \tilde\gamma_{J^{(0)}}\Bigl(\frac{y}{Q^2}\Bigr) +
 \tilde\gamma_{\delta\delta}^S\Bigl(\frac{y}{Q}\Bigr) \biggr] 
  H_2(Q) \tilde J^{(0)}\Bigl(\frac{y}{Q^2} \Bigr) \tilde J^{(0)}\Bigl(\frac{y}{Q^2}\Bigr)  \tilde S_{\delta}^{(2)}\Bigl(\frac{y}{Q}\Bigr) 
 \nn\\
 &\quad + \frac{1}{Q} \gamma_{\delta\theta}^{S}[\alpha_s] 
  H_2(Q) \tilde J^{(0)}\Bigl(\frac{y}{Q^2} \Bigr) \tilde J^{(0)}\Bigl(\frac{y}{Q^2}\Bigr)  \tilde S_{\theta}^{(2)}\Bigl(\frac{y}{Q}\Bigr) 
  \,. \nn
\end{align}
Using the relation between anomalous dimensions that follows from the leading power consistency relation, $\gamma_{H^{(0)}}(Q)+2 \tilde \gamma_{J^{(0)}}(y/Q^2)+\tilde \gamma_{S^{(0)}}(y/Q)=0$, dividing by $\bigl[ \tilde J^{(0)}\bigl(\frac{y}{Q^2} \bigr)\bigr]^2  \tilde S^{(0)}\bigl(\frac{y}{Q}\bigr)$, and multiplying by $iy$ simplifies this result to
\begin{align}
 0 &= 2 H_1(Q) \biggl[ \gamma_{H_1}(Q) -\gamma_{H^{(0)}}(Q) + \tilde\gamma_{\delta\delta}^J\Bigl(\frac{y}{Q^2}\Bigr) -
 \tilde\gamma_{J^{(0)}}\Bigl(\frac{y}{Q^2}\Bigr)  \biggr] 
 \,\Biggl[
 \frac{ \frac{iy}{Q^2} \, \tilde J^{(2)}_{\delta}\!\bigl(\frac{y}{Q^2}\bigr) }
 {\tilde J^{(0)}\bigl(\frac{y}{Q^2} \bigr) } \Biggr]
  \nn \\
 & \quad +   H_2(Q)   \biggl[ \gamma_{H_2}(Q) -\gamma_{H^{(0)}}(Q) + 
 \gamma_{\delta\delta}^S\Bigl(\frac{y}{Q}\Bigr) -
 \tilde\gamma_{S^{(0)}}\Bigl(\frac{y}{Q}\Bigr) \biggr] 
 \,\Biggl[ 
  \frac{\frac{iy}{Q}\,  \tilde S_{\delta}^{(2)}\!\bigl(\frac{y}{Q}\bigr) }
  { \tilde S^{(0)}\bigl(\frac{y}{Q}\bigr) } \Biggr] 
 \nn\\
 &\quad + 2 H_1(Q) \, \gamma_{\delta\theta}^J[\alpha_s] \Biggl[
 \frac{\frac{iy}{Q^2}\,\tilde J^{(2)}_{\theta}\!\bigl(\frac{y}{Q^2}\bigr)}
  { \tilde J^{(0)}\bigl(\frac{y}{Q^2} \bigr) } \Biggr]
  + H_2(Q)\,  \gamma_{\delta\theta}^{S}[\alpha_s] \Biggl[
 \frac{\frac{iy}{Q} \, \tilde S_{\theta}^{(2)}\!\Bigl(\frac{y}{Q}\Bigr)}
 { \tilde S^{(0)}\bigl(\frac{y}{Q}\bigr) } \Biggr] 
  \,. 
\end{align}
This consistency equation is quite non-trivial since it involves separate functions of each of $Q$, $y/Q^2$, and $y/Q$. Specializing to LL order we include only the logarithmic terms from the anomalous dimensions in the first two lines, and only the ${\cal O}(\alpha_s)$ terms for the anomalous dimensions in the last line. This gives
\begin{align} \label{eq:LLconsistency}
 0 &= \Biggl[
 \frac{ \frac{iy}{Q^2} \, \tilde J^{(2)}_{\delta}\!\bigl(\frac{y}{Q^2},\mu\bigr) }
 {\frac{\alpha_s(\mu)}{4\pi}\, \tilde J^{(0)}\bigl(\frac{y}{Q^2},\mu \bigr) } \Biggr]^{\rm LL}
  \frac{\alpha_s^2(\mu)}{(4\pi)^2} 
 \biggl\{ 2 \bigl(\Gamma_{H_1}^0 -\Gamma_{H^{(0)}}^0 \bigr) \log\frac{\mu^2}{Q^2} + 
 2 \bigl( \Gamma_{\delta\delta}^{J0} -  \Gamma^0_{J^{(0)}} \bigr) \log\frac{iy\mu^2}{Q^2}  \biggr\}
  \nn \\
 & +  \biggl[ \frac{ H_2(Q,\mu) }{ H_1(Q,\mu) }\biggr]^{\rm LL} \Biggl[ 
  \frac{\frac{iy}{Q}\,  \tilde S_{\delta}^{(2)}\!\bigl(\frac{y}{Q},\mu\bigr) }
  {\frac{\alpha_s(\mu)}{4\pi}\, \tilde S^{(0)}\bigl(\frac{y}{Q},\mu\bigr) } \Biggr]^{\rm LL} 
 \frac{\alpha_s^2(\mu)}{(4\pi)^2} \biggl\{ \bigl(\Gamma_{H_2}^0 -\Gamma^0_{H^{(0)}} \bigr) \log\frac{\mu^2}{Q^2} + 
 \bigl( \Gamma_{\delta\delta}^{S0} - \Gamma^0_{S^{(0)}} \bigr) 
   \log\frac{i y \mu}{Q} \biggr\}
 \nn\\
 &+   \Biggl[
 \frac{\frac{iy}{Q^2}\,\tilde J^{(2)}_{\theta}\!\bigl(\frac{y}{Q^2},\mu\bigr)}
  { \tilde J^{(0)}\bigl(\frac{y}{Q^2},\mu\bigr) } \Biggr]^{\rm LL} 
  \, 2 \frac{\alpha_s(\mu)}{4\pi}\, \gamma_{\delta\theta}^{J0} 
  +  \biggl[\frac{ H_2(Q,\mu) }{ H_1(Q,\mu) }\biggr]^{\rm LL}  \Biggl[
 \frac{\frac{iy}{Q} \, \tilde S_{\theta}^{(2)}\!\Bigl(\frac{y}{Q},\mu\Bigr)}
 { \tilde S^{(0)}\bigl(\frac{y}{Q},\mu\bigr) } \Biggr]^{\rm LL} 
 \, \frac{\alpha_s(\mu)}{4\pi}\, \gamma_{\delta\theta}^{S0} 
  \,, 
\end{align}
where we have restored the $\mu$ arguments. The $0$ superscripts on the anomalous dimensions here indicate that these are the lowest order term in these anomalous dimensions (which are simple numbers). In the first two lines we have included a $1/\alpha_s(\mu)$ since $\tilde J_\delta^{(2)}$ and $\tilde S_\delta^{(2)}$ themselves start at ${\cal O}(\alpha_s)$. This way all terms in square brackets in \Eq{eq:LLconsistency} start at ${\cal O}(\alpha_s^0)$.  Since $\mu$ is  arbitrary, all ratios of hard, jet, and soft functions in square brackets in \Eq{eq:LLconsistency} can each be thought of as a LL series, $\big[\cdots\big]^{\rm LL} =  \sum_{k=0}^\infty a_k [\alpha_s(\mu) \log^2(X)]^k$, where $X=\mu^2/Q^2$, $X=y\mu^2/Q^2$, or $X=y\mu/Q$ for ratios of hard, jet, or soft functions respectively (or the analogs with running coupling effects which does not change the arguments below). The coefficients $a_k$ in these series are numbers that depend on powers of the corresponding anomalous dimensions for the objects in that square bracket. 

To see what \Eq{eq:LLconsistency} implies, first consider the ratio of jet functions in the first line. In the case of our illustrative example from \Sec{sec:illexample} we have $\tilde J_\delta^{(2)}/J^{(0)} \propto d/d(y/Q^2) \, \log \tilde J^{(0)}$, so it is safe to assume that this ratio of jet functions is a non-trivial function of $y/Q^2$. The first line of \Eq{eq:LLconsistency} can then not cancel against the terms in the second line since they have different functional dependence on $y$ and $\mu/Q$. Nor can it cancel against the terms on the third line, since they start at different orders in $\alpha_s$.  This implies that the curly bracket on the first line of \Eq{eq:LLconsistency} vanishes. Due to the presence of two independent types of logarithms in this bracket this immediately implies relations between the cusp anomalous dimension coefficients for these functions at LL order:
\begin{align} \label{eq:consistencyJ}
  \Gamma_{H_1}^0 &= \Gamma_{H^{(0)}} \,, 
 & \Gamma_{\delta\delta}^{J0} &=  \Gamma_{J^{(0)}} \,.
\end{align}
For the same reason the curly bracket on the second line of \Eq{eq:LLconsistency} must also vanish, which then implies the following LL anomalous dimension relations:
\begin{align} \label{eq:consistencyS}
  \Gamma_{H_2}^0 &= \Gamma_{H^{(0)}} \,, 
 & \Gamma_{\delta\delta}^{S0} &=  \Gamma_{S^{(0)}} \,.
\end{align}
Together these imply that $\Gamma_{H_1}^0=\Gamma_{H_2}^0$, which gives $[H_2(Q,\mu)/H_1(Q,\mu)]^{\rm LL}=1$. 

In \Eq{eq:LLconsistency} this then leaves only the LL mixing terms, where the remaining constraint now takes the form
\begin{align}
 0=  \Biggl[
 \frac{\frac{iy}{Q^2}\,\tilde J^{(2)}_{\theta}\!\bigl(\frac{y}{Q^2},\mu\bigr)}
  { \tilde J^{(0)}\bigl(\frac{y}{Q^2},\mu\bigr) } \Biggr]^{\rm LL} 
  \, 2 \, \gamma_{\delta\theta}^{J0} 
  +   \Biggl[
 \frac{\frac{iy}{Q} \, \tilde S_{\theta}^{(2)}\!\Bigl(\frac{y}{Q},\mu\Bigr)}
 { \tilde S^{(0)}\bigl(\frac{y}{Q},\mu\bigr) } \Biggr]^{\rm LL} 
 \,  \gamma_{\delta\theta}^{S0} 
 \,.
\end{align}
In our illustrative example the two square brackets here are both equal to $1$.  The RG consistency implies that this is actually a much more general result, true for any operators satisfying the assumptions set out at the beginning of this section. In particular, since the two square brackets have different functional dependence, $y/Q^2$ and $y/Q$ respectively, they must both be independent of these variables. This gives:\footnote{More generally the RHS of the results in \Eq{eq:thetaconsistency} could be constants, but we choose to normalize $\tilde J^{(2)}_{\theta}$ and $\tilde S_{\theta}^{(2)}$ so these constants are both $1$.}
\begin{align} \label{eq:thetaconsistency}
 & \Biggl[
 \frac{\frac{iy}{Q^2}\,\tilde J^{(2)}_{\theta}\!\bigl(\frac{y}{Q^2},\mu\bigr)}
  { \tilde J^{(0)}\bigl(\frac{y}{Q^2},\mu\bigr) } \Biggr]^{\rm LL} = 1 \,,
 & \Biggl[
 \frac{\frac{iy}{Q} \, \tilde S_{\theta}^{(2)}\!\Bigl(\frac{y}{Q},\mu\Bigr)}
 { \tilde S^{(0)}\bigl(\frac{y}{Q},\mu\bigr) } \Biggr]^{\rm LL} & =1 \,.
\end{align}
This then leaves a simple relation between the mixing anomalous dimensions
\begin{align} \label{eq:gamthetaconsitency}
  2\gamma_{\delta \theta}^{J0} + \gamma_{\delta\theta}^{S0} = 0 \,,
\end{align}
which we also found in our illustrative example. 
In momentum space \Eq{eq:thetaconsistency} implies that
\begin{align} \label{eq:JSthetaconstraint}
  J_\theta^{(2)}(s,\mu)^{\rm LL} &= \int_0^s \! ds'\, J^{(0)}(s',\mu)^{\rm LL} 
   \,,
 & S_\theta^{(2)}(k,\mu)^{\rm LL} &= \int_0^k \! dk'\, S^{(0)}(k',\mu)^{\rm LL} 
   \,.
\end{align}
While true in our illustrative example, viewed as a more general constraint  this result is quite interesting. For more general operators defining $J_\delta^{(2)}$ and $S_\delta^{(2)}$ it might not be a priori clear (without performing higher order loop and gluon emission calculations) how to define the operators giving the $J_\theta^{(2)}$ and $S_\theta^{(2)}$ that one mixes into. The RG consistency result in \Eq{eq:JSthetaconstraint}  implies that the required $J_\theta^{(2)}$ and $S_\theta^{(2)}$ functions agree with those defined from the cumulative of the leading power operators, at least at LL order.   The RG consistency results in \Eqs{eq:consistencyJ}{eq:consistencyS} furthermore imply that the LL cusp anomalous dimensions of $J_\delta^{(2)}$ and $S_\delta^{(2)}$ are the same as those for the jet and soft functions at leading power. Note that although $\gamma_{\theta\theta}^J$ or $\gamma_{\theta\theta}^S$ do not appear explicitly in the RG consistency equation, they are present in the LL expressions for $\tilde J^{(2)}_{\theta}$ and $\tilde S^{(2)}_{\theta}$ and hence are constrained by \Eq{eq:thetaconsistency}.

This example also illustrates another important point. There must always be (at least) a pair of functions at subleading power whose renormalization group evolution is tied by consistency. This is also clear from the fact that when evaluated at their natural scales, the subleading power $J_\delta^{(2)}$ and $S_\delta^{(2)}$ functions are $0+\cO(\alpha_s)$, and not $\delta(\tau)+\cO(\alpha_s)$ as at leading power. Thus if one chooses to run all functions to the canonical scale of  either of the subleading power functions, this function will simply not contribute at LL accuracy. To see this explicitly, we can use the evolution equations to run all functions in the position space factorization theorem from their canonical scales $\mu_H^2\sim Q^2$, $\mu_J^2\sim Q^2/iy$, or $\mu_S^2\sim Q^2/(iy)^2$ to a common scale $\mu^2$.  This gives
\begin{align}
  \frac{1}{\sigma_0} \frac{\df\sigma^{(2)}}{\df y}
  &= \frac{2}{Q^2} H_1(Q,\mu_H) U_{H_1}(Q,\mu_H,\mu) 
  U_{J}^{(0)}\Bigl(\frac{y}{Q^2},\mu_J,\mu\Bigr) 
  \tilde J^{(0)}\Bigl(\frac{y}{Q^2},\mu_J \Bigr)
  U_{S}^{(0)}\Bigl(\frac{y}{Q},\mu_S,\mu\Bigr) 
 \nn\\
 &\quad\ \times
 \tilde S^{(0)}\Bigl(\frac{y}{Q},\mu_S\Bigr)  
 \biggl[ U_{\delta\delta}^J\Bigl(\frac{y}{Q^2},\mu_J,\mu\Bigr) \tilde J^{(2)}_{\delta}\Bigl(\frac{y}{Q^2},\mu_J \Bigr) +
  U_{\delta\theta}^J\Bigl(\frac{y}{Q^2},\mu_J,\mu\Bigr) \tilde J^{(2)}_{\theta}\Bigl(\frac{y}{Q^2},\mu_J \Bigr) \biggr]
   \nn  \\
   &\ \
  + \frac{1}{Q} H_2(Q,\mu) U_{H_2}(Q,\mu_H,\mu) 
  \Bigl[ U_J^{(0)}\Bigl(\frac{y}{Q^2},\mu_J,\mu \Bigr) \tilde J^{(0)}\Bigl(\frac{y}{Q^2},\mu_J \Bigr) \Bigr]^2 
  \nn\\
  & \quad\ \times 
  \biggl[ U_{\delta\delta}^{S}\Bigl(\frac{y}{Q},\mu_S,\mu\Bigr) 
   \tilde S_{\delta}^{(2)}\Bigl(\frac{y}{Q},\mu_S\Bigr)  + 
 U_{\delta\theta}^{S}\Bigl(\frac{y}{Q},\mu_S,\mu\Bigr) 
   \tilde S_{\theta}^{(2)}\Bigl(\frac{y}{Q},\mu_S\Bigr)  \biggr]
  \,.
\end{align}
Here the $U_H$, $U_S$ and $U_J$ factors are evolution kernels for the various hard, jet, and soft functions. For our analysis of $H\to gg$ in pure glue QCD  their explicit form will be given later in the text.
At LL order we can then use that
\begin{align}
 \tilde J^{(2)}_{\delta}(y/Q^2,\mu_J )  &=0+\cO(\alpha_s)\,,
 & \tilde S^{(2)}_{\delta}(y/Q,\mu_S ) &=0+\cO(\alpha_s)\,,
\end{align}
which implies that the terms with the $U_{\delta\delta}^J$ and $U_{\delta\delta}^S$ kernels are not needed at this order. We can also simplify the LL result by using $\tilde S^{(0)} = 1$ and $\tilde J^{(0)}=1$ (we allow here a non-trivial overall numeric factor from $H_1$ and $H_2$ at tree level). 
The LL resummed result then simplifies to 
\begin{align} \label{eq:LLresum_arb_mu}
  \frac{1}{\sigma_0} \frac{\df\sigma^{(2)\,{\rm LL}}}{\df y}
  &= \frac{2 H_1}{Q^2} U_{H_1}(Q,\mu_H,\mu) 
  U_{J}^{(0)}\Bigl(\frac{y}{Q^2},\mu_J,\mu\Bigr) 
  U_{S}^{(0)}\Bigl(\frac{y}{Q},\mu_S,\mu\Bigr) 
 U_{\delta\theta}^J\Bigl(\frac{y}{Q^2},\mu_J,\mu \Bigr) \tilde J^{(2)}_{\theta}\Bigl(\frac{y}{Q^2},\mu_J \Bigr) 
   \nn  \\
   &\ \
  + \frac{H_2}{Q}  U_{H_2}(Q,\mu_H,\mu) 
  \Bigl[ U_J^{(0)}\Bigl(\frac{y}{Q^2},\mu_J,\mu \Bigr)  \Bigr]^2   
  U_{\delta\theta}^{S}\Bigl(\frac{y}{Q},\mu_S,\mu\Bigr) 
   \tilde S_{\theta}^{(2)}\Bigl(\frac{y}{Q},\mu_S\Bigr)  
  \,.
\end{align}
Finally we can use the RG consistency freedom that says the same result is obtained no matter what value we pick for $\mu$. For example, taking $\mu=\mu_J$ we have $U_{\delta \theta}^J(y/Q^2,\mu_J,\mu_J)=0$ which removes the first term, and $U_{J}^{(0)}(y/Q^2,\mu_J,\mu_J)=1$ which simplifies the second, leaving
\begin{align}  \label{eq:LLresum_pick_muJ}
  \frac{1}{\sigma_0} \frac{\df\sigma^{(2)\,{\rm LL}}}{\df y}
  &=  \frac{H_2}{Q}  U_{H_2}(Q,\mu_H,\mu_J)  
  U_{\delta\theta}^{S}\Bigl(\frac{y}{Q},\mu_S,\mu_J\Bigr) 
   \tilde S_{\theta}^{(2)}\Bigl(\frac{y}{Q},\mu_S\Bigr)  
  \,.
\end{align}
In this form the  LL resummed result is obtained completely from the subleading power soft function. If instead we had chosen $\mu=\mu_S$, then $U_{\delta\theta}^S(y/Q,\mu_S,\mu_S)=0$ would have removed the second term in \Eq{eq:LLresum_arb_mu}, and the result would have been expressed entirely from the first term that involves the subleading power jet functions, which can be simplified using $U_S^{(0)}(y/Q,\mu_S,\mu_S)=1$.  This equivalence between different resummed formula is an expression of the LL consistency result in \Eq{eq:gamthetaconsitency} at the level of the cross section. We will use \Eq{eq:LLresum_pick_muJ} to simplify the resummation for thrust at next-to-leading power in \Sec{sec:fact}.

\section{Solution to the Subleading Power RG Mixing Equation}\label{sec:solution}

Having illustrated that the renormalization of subleading power jet and soft functions generically involves mixing with $\theta$-jet and $\theta$-soft operators, in this section we solve a general form of the subleading power RG equations involving mixing, including the running coupling $\alpha_s(\mu)$. This solution will be sufficient for all cases required in this paper, and we believe that it will be of general utility for subleading power resummation.

We consider a function, $F$, which obeys an RG equation of the form of \Eq{eq:2by2mix}.  To remove the convolution structure, we work in Fourier (or Laplace) space, with a variable $y$ conjugate to a momentum variable $k$ of dimension $p$. Defining
\begin{align}
\tilde F(y)= \int dk~ e^{-ik y}~ F(k),
\end{align}
the RG equation for $\tilde F$ is then multiplicative
\be\label{eq:RG_to_solve}
	 \mdm \columnspinor{\tilde F^{(2)}_{\delta} (y,\mu)}{\tilde F^{(2)}_{\theta}(y,\mu)} = \smallmatrix{\tilde \gamma_{11}(y,\mu)}{ \gamma_{12}[\alpha_s]}{0}{\tilde \gamma_{22}(y,\mu)} \columnspinor{\tilde F^{(2)}_{\delta}(y,\mu)}{\tilde F^{(2)}_{\theta}(y,\mu)} \,.
\ee
Here, to simplify notation, we have defined
\begin{align}
\tilde \gamma_{11} (y,\mu) &=\Gamma_{11}[\alpha_s] \log \bigl(ie^{\gamma_E} (y-i0)\mu^p  \bigr)+\gamma_{11}[\alpha_s]\,, \\
\tilde \gamma_{22} (y,\mu) &=\Gamma_{22}[\alpha_s] \log \bigl(ie^{\gamma_E} (y-i0)\mu^p \bigr) +\gamma_{22}[\alpha_s]\,.\nn
\end{align}
To shorten the equations, we will not explicitly write the branch cut prescription in the following. The off-diagonal mixing term, $\gamma_{12}[\alpha_s]$, does not contain logarithms.

\subsection{General Solution}

We will solve the subleading power mixing equation without the constraint that $\tilde \gamma_{11}=\tilde\gamma_{22}$, as occurred in the example of \Sec{sec:renorm_a}. We do this both because we believe that this solution will be relevant for the renormalization of more general functions at subleading power, as well as to illustrate how the standard leading power Sudakov exponential arises as a special limit when $\tilde \gamma_{11}= \tilde \gamma_{22}$, but not more generally.

We can write the all orders solution to the differential equation of \Eq{eq:RG_to_solve} as
\begin{align}\label{eq:solution_gen}
\tilde F^{(2)}_{ \delta} (y,\mu)
 = U_{\delta\delta}(y,\mu,\mu_0)\,  \tilde F^{(2)}_{\delta}(y,\mu_0) 
  + U_{\delta\theta}(y,\mu,\mu_0)\, \tilde F^{(2)}_{\theta}(y,\mu_0)  \,,
\end{align}
with
\begin{align}\label{eq:solution_gen_b}
 U_{\delta\delta}(y,\mu,\mu_0) &= \exp\biggl[\: \int\limits_{\mu_0}^\mu \dmmp \,\tilde \gamma_{11}(y,\mu') \biggr] \,,
 & U_{\delta\theta}(y,\mu,\mu_0) &= U_{\delta\delta}(y,\mu,\mu_0) X(y,\mu, \mu_0) \,,
\end{align}
where $X$ satisfies
\begin{align}
\mdm X(y,\mu, \mu_0) =e^{-\int\limits_{\mu_0}^\mu \dmmp \tilde \gamma_{11}(y,\mu') }  \gamma_{12}[\alpha_s(\mu)] \,   e^{~\int\limits_{\mu_0}^\mu \dmmp \tilde \gamma_{22}(y,\mu')} \,,
\end{align}
and the boundary condition $X(y,\mu_0,\mu_0)=0$. 
Solving for $X$, we have
\begin{align}\label{eq:solution_X_first}
X(y,\mu, \mu_0) &=\int\limits_{\mu_0}^\mu \frac{d\mu^{''}}{\mu^{''}} e^{-\int\limits_{\mu_0}^{\mu''} \dmmp \tilde \gamma_{11}(y,\mu') }  \gamma_{12}[\alpha_s(\mu '')] \,   e^{~\int\limits_{\mu_0}^{\mu''} \dmmp \tilde \gamma_{22}(y,\mu')} \\
&=  \int\limits_{\mu_0}^\mu \frac{d\mu^{''}}{\mu^{''}} \: \gamma_{12}  [\alpha_s(\mu'')]\: \exp\Biggl( {-\int\limits_{\mu_0}^{\mu''} \dmmp [ \tilde \gamma_{11}(y,\mu') -\tilde \gamma_{22}(y,\mu') ] }  \Biggr)
 \,.\nn
\end{align}
We can derive a closed analytic form for $X$ order by order in the anomalous dimensions, including the running coupling. For the remainder of this section we consider the solution at LL order, where the anomalous dimensions take the form 
\be\label{eq:LLapproxconditions}
\tilde \gamma_{ii}(y,\mu) 
  =\Gamma_{ii}^0 \, \frac{\alpha_s(\mu)}{4\pi}\, \log \left (\frac{\mu^p}{\mu_y^p} \right ) 
 \,,\qquad \gamma_{12}[\alpha_s]
  =\gamma_{12}^0\, \frac{\alpha_s(\mu)}{4\pi} \,,
\ee
where $\Gamma^0_{11}$, $\Gamma^0_{22}$, $\gamma^0_{12}$ are numbers, and  we have defined the mass dimension $1$ variable $\mu_y$ by
\be\label{eq:muydef}
	\frac{1}{\mu^p_y} \equiv e^{\gamma_E}i(y-i0) \,.
\ee
Note that at LL order we need only the logarithmic term for the diagonal anomalous dimensions $\tilde\gamma_{11}(y,\mu)$ and $\tilde\gamma_{22}(y,\mu)$. The non-logarithmic term is needed for the off-diagonal term $\gamma_{12}[\alpha_s]$ because of the fact that the boundary terms in \Eq{eq:solution_gen} start at different orders, $\tilde F^{(2)}_{\delta}(y,\mu_0)\sim {\cal O}(\alpha_s)$ and $\tilde F^{(2)}_{\theta}(y,\mu_0) \sim {\cal O}(\alpha_s^0)$. 

To include the effects of running coupling, we use the standard approach of switching to an integration in $\alpha_s$ instead of $\mu$ through the change of variables
\begin{align}
	\frac{d\mu}{\mu}= \frac{d\alpha_s}{\beta[\alpha_s]}\,.
\end{align}
At LL-order, we can use the LL $\beta$ function which gives
\begin{align}\label{eq:changeofvariable}
	\frac{d\mu}{\mu}= -\frac{2\pi}{\beta_0}\frac{d\alpha_s}{\alpha_s^2}\,, \qquad \beta_0 = \frac{11}{3} C_A - \frac{4}{3}T_F n_f\,,
\end{align}
We also rewrite the logarithm appearing in the anomalous dimension as
\be\label{eq:logmuasalpha}
	\log\Bigl( \frac{\mu}{\mu_y}\Bigr) = -\frac{2\pi}{\beta_0}\int_{\alpha_s(\mu_y)}^{\alpha_s(\mu)} \daap = \frac{2\pi}{\beta_0}\left(\frac{1}{\alpha_s(\mu)}-\frac{1}{\alpha_s(\mu_y)}\right) = \frac{2\pi}{\beta_0 \alpha_s(\mu_y)}\left(\frac{\alpha_s(\mu_y)}{\alpha_s(\mu)}-1\right)\,.
\ee
We then have
\begin{align}  \label{eq:Uddresult}
	U_{\delta\delta}(y,\mu,\mu_0) &= \exp\Biggl\{~\Gamma_{11}^0\int\limits_{\mu_0}^\mu \dmmp \left(\frac{\alpha_s(\mu')}{4\pi}\right)\log \left (\frac{\mu^{'p}}{\mu^p_y} \right )\Biggr\}  \\
	&=\exp\left[\frac{p\pi\Gamma^0_{11}}{\beta_0^2\alpha_s(\mu_0)} \left(\frac{1}{r} - 1 + \log r \right) \right]\left(\frac{\mu^p_y}{\mu^p_0}\right)^{\frac{\Gamma^0_{11}}{2\beta_0}\log(r) }\,, \nn
\end{align}
where 
\be\label{eq:rdefinition}
	r\equiv\frac{\alpha_s(\mu)}{\alpha_s(\mu_0)} \,,
\ee 
and at this order we take the boundary conditions
 \begin{align}
 \tilde F^{(2)}_{\delta}(y,\mu_0)=0\,, \qquad \tilde F^{(2)}_{\theta}(y,\mu_0)=\frac{1}{i(y-i0)}\,.
 \end{align}
Recall that $1/i(y-i0)$ is the Fourier transform of $\theta(k)$.
Thus at LL the solution becomes
\begin{align}\label{eq:solution_gen_LL}
	\tilde F^{(2) \LL}_{ \delta} (y,\mu) = U_{\delta\theta}(y,\mu,\mu_0)  \, \frac{1}{i(y-i0)} \,,
\end{align}
with the evolution kernel given by 
\begin{align} \label{eq:ULLrunning}
	U_{\delta\theta}^{\LL}(y,\mu,\mu_0) &=\exp\left[\frac{p\pi\Gamma^0_{11}}{\beta_0^2\alpha_s(\mu_0)} \left(\frac{1}{r} - 1 + \log r \right) \right]\left(\frac{\mu^p_y}{\mu^p_0}\right)^{\frac{\Gamma^0_{11}}{2\beta_0}\log(r) }X^{\LL}(y,\mu, \mu_0) \,.
\end{align}
Using \Eqs{eq:changeofvariable}{eq:logmuasalpha} we can compute $X(y,\mu, \mu_0)$ in terms of the running coupling as
\begin{align}\label{eq:Xmostgeneral}
	\!X(y,\mu, \mu_0) &= -\frac{\gamma^0_{12}}{2\beta_0} \int_{\alpha_s(\mu_0)}^{\alpha_s(\mu)} \frac{\df \alpha_s'}{\alpha_s'} \exp\left\{\!\frac{p \pi }{\beta_0^2}\Bigl(\Gamma^0_{11} -\Gamma^0_{22}\Bigr)\! \int_{\alpha_s(\mu_0)}^{\alpha_s'} \frac{\df \alpha_s''}{\alpha_s''}\left(\frac{1}{\alpha_s''}-\frac{1}{\alpha_s(\mu_y)}\right) \!\right\} 
     \nn \\
 &= -\frac{\gamma^0_{12}}{2\beta_0} \int_{\alpha_s(\mu_0)}^{\alpha_s(\mu)} \frac{\df \alpha_s'}{\alpha_s'} \exp\left\{ \frac{p \pi}{\beta_0^2}\left(\Gamma^0_{11} -\Gamma^0_{22}\right) \left[\frac{1}{\alpha_s(\mu_0)} - \frac{1}{\alpha_s'} - \frac{1}{\alpha_s(\mu_y)} \log\frac{\alpha_s'}{\alpha_s(\mu_0)} \right]\right\}
     \nn \\
 &= -\frac{\gamma^0_{12}}{2\beta_0} \int_{\phi(\mu)}^{\phi(\mu_0)} 
  \frac{\df \phi'}{\phi'} \exp\left\{ \phi(\mu_0) - \phi' -\phi(\mu_y) \log\frac{\phi(\mu_0)}{\phi'} \right\}
  \,,
\end{align}
where in the last line we used the definition
\begin{align}
	\phi(\mu)\equiv \frac{p\pi(\Gamma^0_{11}-\Gamma^0_{22})}{\beta_0^2\,\alpha_s(\mu)}\,.
\end{align}
The final integral gives the LL solution
\begin{align} \label{eq:XLLresult}
	X^\LL(y,\mu, \mu_0)
	&=-\frac{\gamma^0_{12}}{2\beta_0} e^{\phi(\mu_0)}\left[ r^{-\phi(\mu_y)} E\Bigl(1-\phi(\mu_y),\phi(\mu)\Bigr)-E\Bigl(1-\phi(\mu_y),\phi(\mu_0)\Bigr) \right] ,
\end{align}
where $E(n,z)$ is the exponential integral function 
\begin{align}
 E(n,z)=\int_1^{\infty }  \! \frac{dt}{t^n} \: e^{-zt}\,.
\end{align}
Plugging these results into \Eq{eq:solution_gen} we obtain the general solution to the subleading RG at LL order in terms of the results in \Eqs{eq:Uddresult}{eq:XLLresult}:
\begin{align} \label{eq:solution_LL}
\tilde F^{(2)}_{ \delta} (y,\mu)^{\rm LL}
 =  U_{\delta\delta}^{\rm LL}(y,\mu,\mu_0)\, X^{\rm LL}(y,\mu,\mu_0)\,
     \frac{1}{i(y-i0)}  \,.
\end{align}

For illustration we can take the limit 
without the running coupling, set $\mu_0 = \mu_y$, and assume\footnote{Note that we made no assumption on the signs of the $\Gamma_{11}^0$ and $\Gamma_{22}^0$ which can be negative. If $\Gamma_{11}^0<\Gamma_{22}^0$, the result involves an imaginary error function ($\Erfi$) instead of the error function ($\Erf$).} $\Gamma_{11}^0>\Gamma_{22}^0$ which gives  
\begin{align}\label{eq:ULLfixedalphas}
	U_{\delta\theta}^{LL}(y,\mu,\mu_y)\biggr|_{\alpha_s(\mu) = \alpha_s}
   &
      =\gamma^0_{12} \,\frac{\alpha_s}{8\pi} \,
    \sqrt{\frac{\pi}{\Delta_\Gamma}}\,\Erf\biggl[ 
    \sqrt{\Delta_{\Gamma}} \,  \log\frac{\mu}{\mu_y} \biggr]\,
    \exp\left[ p \, \Gamma^0_{11}\, \frac{\alpha_s}{8\pi}\, \log^2\frac{\mu}{\mu_y}  \right]  ,
\end{align}
where $\Delta_\Gamma \equiv\left(\frac{\alpha_s}{8\pi}\right)p\left(\Gamma^0_{11}-\Gamma^0_{22}\right)$ and Erf is the error function, $\Erf(x)=(2/\sqrt{\pi}) \int_0^x e^{-t^2} dt$ which expanded around $x=0$ reads $\Erf(x)=2x/\sqrt{\pi}-2x^3/(3 \sqrt{\pi })+\cO(x^5)$. The kernel in \Eq{eq:ULLfixedalphas} is easily interpreted as the standard Sudakov factor with fixed coupling multiplied by the error function arising from the integral over the difference of Sudakov exponentials in \Eq{eq:solution_X_first}.  The solutions in \Eqs{eq:ULLrunning}{eq:ULLfixedalphas} emphasize that there is a closed form solution in terms of elementary functions, and that in the most general case we will not necessarily get a simple Sudakov exponential at subleading power. We also emphasize that in all the LL results $\gamma^0_{12}$ appears only as an overall factor.

\subsection{Solution With Equal Diagonal Entries}
To gain further insight into the form of the LL solution to the subleading power RG it is instructive to restrict our attention to the case $\Gamma^0_{11}=\Gamma^0_{22}$ which is the relevant one for the subleading soft and jet functions considered in \Sec{sec:renorm}. With $\Gamma^0_{11}=\Gamma^0_{22}$, we have $\phi=0$ so that $X$ simplifies to 
\begin{align}
X^\LL(y,\mu,\mu_0)\big|_{\Gamma^0_{11}
 =\Gamma^0_{22}}&=-\frac{\gamma^0_{12}}{2\beta_0} \int_{\alpha_s(\mu_0)}^{\alpha_s(\mu)} \frac{\df \alpha_s'}{\alpha_s'} = -\frac{\gamma^0_{12}}{2\beta_0}\log r \,,
\end{align}
where $r$ was defined in \Eq{eq:rdefinition} and the evolution kernel simplifies to
\begin{align} \label{eq:ULLrunningsameGamma}
	U_{\delta\theta}^{\LL}(y,\mu,\mu_0)\big|_{\Gamma^0_{11}=\Gamma^0_{22}} &=-\frac{\gamma^0_{12}}{2\beta_0}\, 
   \log r\,\exp\left[\frac{p\pi\Gamma^0_{11}}{\beta_0^2\alpha_s(\mu_0)} \left(\frac{1}{r} - 1 + \log r \right) \right]\left(\frac{\mu^p_y}{\mu^p_0}\right)^{\frac{\Gamma^0_{11}}{2\beta_0}\log(r) } \,.
\end{align}
Therefore with $\Gamma_{11}^0=\Gamma_{22}^0$ we recover a simple Sudakov evolution at LL. For this case the final expression for the LL resummed function in position space is
\be\label{eq:FLL}
	\tilde F^{(2)\LL}_{\delta}(y,\mu)
  = -\frac{\gamma^0_{12}}{2\beta_0}\log r\exp\left[\frac{p\pi\Gamma^0_{11}}{\beta_0^2\alpha_s(\mu_0)} \left(\frac{1}{r} - 1 + \log r \right) \right]\left(\frac{\mu^p_y}{\mu^p_0}\right)^{\frac{\Gamma^0_{11}}{2\beta_0}\log(r) } \frac{1}{i(y-i0)}\,.
\ee

To obtain the expression for $F^{(2)\LL}_{\delta}(k,\mu)$ we  transform \Eq{eq:FLL} back to momentum space  which gives
\begin{align}\label{eq:Fmomspacetext}
	F^{(2)\LL}_{\delta}(k,\mu)
  &= 	U^{\LL}_{\delta\theta}(k,\mu,\mu_0) \, \theta(k)  \,, 
\end{align}
where the evolution kernel is obtained with the simple replacement $\mu^p_y\to k$,
\begin{align}\label{eq:Umomspace}
U^{\LL}_{\delta\theta}(k,\mu,\mu_0) &= - \frac{\gamma^0_{12}}{2\beta_0}\log r\exp\left[\frac{p\pi\Gamma^0_{11}}{\beta_0^2\alpha_s(\mu_0)} \left(\frac{1}{r} - 1 + \log r \right) \right] \left(\frac{k}{\mu^p_0}\right)^{\frac{\Gamma^0_{11}}{2\beta_0}\log(r) }  \,.
\end{align}
Further details about why this simple replacement suffices at LL are given in \App{sec:inversefourier}.

For concreteness, let us now consider the case where the subleading function $F^{(2)}_\delta(k,\mu)$ is the subleading power soft function of \Eq{eq:tau_funcs}. The soft function depends on a momentum variable of dimension $p=1$ and from \Eqs{eq:anom_dim_mix_diag}{eq:anom_dim_mix} we have that for $S^{(2)}_{g,\delta}(k,\mu)$ the anomalous dimensions are\footnote{The minus sign for $\Gamma^0_{11}$ comes from the fact that Laplace transforming \Eq{eq:LP_anom_dim} we have 
\be 
	\frac{1}{\mu} \biggl[  \frac{\mu\,\theta(k)}{k}  \biggr]_+ \to -\log(ye^{\gamma_E} \mu)\,,  \nn
\ee 
therefore giving 
\be
	\underbrace{-4 \Gamma_{\text{cusp}}^{g,0}}_{\Gamma^0_{11}} \frac{\alpha_s}{4\pi}\log(ye^{\gamma_E} \mu)\,. \nn
\ee }
\begin{align}
	\tilde \gamma_{11} (k,\mu) = \tilde \gamma_{22} (k,\mu) &= \gamma^{S}_{g} (k,\mu) &&\implies&& \Gamma^0_{11} = \Gamma^0_{22}= -4 \Gamma^{g,0}_\cusp = -16 C_A\,, \\
	\gamma_{12} [\alpha_s] &= 4\Gamma^g_\cusp [\alpha_s] &&\implies && \gamma_{12}^0 = 4 \Gamma^{g,0}_\cusp = 16 C_A\,. \nn
\end{align}
Using these results in \Eq{eq:Umomspace} we obtain
\begin{align}\label{eq:Smomspacetext}
	S^{(2)\LL}_{g,\delta}(k,\mu) &= -\theta(k) \frac{2 \Gamma^{g,0}_\cusp}{\beta_0}\log r\exp\left[-\frac{4\pi \Gamma^{g,0}_\cusp}{\beta_0^2\alpha_s(\mu_0)} \left(\frac{1}{r} - 1 + \log r \right) \right] \left(\frac{k}{\mu_0}\right)^{\frac{-2 \Gamma^{g,0}_\cusp}{\beta_0}\log(r) }  \,.
\end{align}
We can resum logarithms in the subleading power soft function by running from the canonical scale of the soft function $\mu_0=\mu_S=Q \tau$, to an arbitrary scale $\mu$. Hence,
\begin{align}\label{eq:SresummedQt}
	S^{(2)\LL}_{g,\delta}(Q\tau,\mu) &= -\theta(\tau) \frac{2 \Gamma^{g,0}_\cusp}{\beta_0}\log \left(\frac{\alpha_s(\mu)}{\alpha_s(Q\tau)}\right)   \\
	&\ \ \times\exp\left[-\frac{4\pi \Gamma^{g,0}_\cusp}{\beta_0^2 \alpha_s(Q\tau)} \left(\frac{\alpha_s(Q\tau)}{\alpha_s(\mu)} - 1 + \log  \frac{\alpha_s(\mu)}{\alpha_s(Q\tau)}\right) \right]
	 . \nn
\end{align}
If we ignore the running of the coupling, this simplifies to
\be\label{eq:SLLfixedcoupling}
	 S^{(2)\LL}_{g,\delta}(Q\tau,\mu)\bigg|_{\alpha_s(\mu)=\alpha_s} \!\!\!\!\!\! = \theta(\tau)4 \Gamma^{g,0}_\cusp\left(\frac{ \alpha_s}{4\pi}\right)\log \left(\frac{\mu}{Q\tau}\right)\exp\left[-2 \Gamma^{g,0}_\cusp\left(\frac{\alpha_s}{4\pi}\right)\log^2 \left(\frac{\mu}{Q\tau}\right)\right] ,
\ee
where the physical interpretation is quite clear. Expanding this structure perturbatively in $\alpha_s$, we have
\begin{align}
S^{(2)}_{g, \delta} (Q\tau,\mu)\bigg|_{\alpha_s(\mu)=\alpha_s} \!\!\!\!\!\!= \theta(\tau) \left[ \left(\frac{ \alpha_s}{4\pi}\right)\gamma^0_{12} \log\left( \frac{\mu}{Q \tau} \right) +  \frac{1}{2} \left(\frac{ \alpha_s}{4\pi}\right)^2 \gamma^0_{12} \Gamma^0_{11} \log^3\left( \frac{\mu}{Q\tau} \right) +\cdots    \right] .
\end{align}
We see that the first single logarithm is generated by the mixing into the $\theta$-function operators, and then this is dressed by a double logarithmic Sudakov that is driven by the diagonal entries in the mixing matrix, namely the cusp anomalous dimensions. This shows again how the single log appearing in the fixed order expansion is generated through RG evolution, namely through operator mixing. Therefore, as desired, all large logarithms are generated through RG evolution, and they are resummed to all orders by solving the subleading power RG equation with mixing. We also see that the operator mixing is absolutely crucial, since the entire LL result comes from the mixing which starts the evolution. 

For completeness, we present also the result for the subleading jet function after LL evolution. The anomalous dimensions are derived in \Eqs{eq:anom_dim_mix_diag}{eq:anom_dim_mix} and are related to the soft function ones via RG consistency. 
\begin{align}
	\tilde \gamma_{11} (k,\mu) = \tilde \gamma_{22} (k,\mu) &= \gamma^{J}_{g} (k,\mu) &&\implies&& \Gamma^0_{11} = \Gamma^0_{22}= 2 \Gamma^{g,0}_\cusp = 8 C_A\,,
\nn\\
    \gamma_{12}[\alpha_s] &= -\frac{1}{2} \gamma_{\delta\theta}^{S0} 
    &&\implies&& \gamma^0_{12} = -2 \Gamma^{g,0}_\cusp = -8 C_A \,.
\end{align}
The canonical scales for $J^{(2)}_{g, \delta} (s,\mu)$ are given by
\begin{align}
s = \mu_J^2 = \mu_0^2 = Q^2\tau \,\quad\implies p = 2\,.
\end{align}
Therefore, we find 
\begin{align}
	J^{(2)}_{g, \delta} (Q^2\tau,\mu) &= \theta(\tau) \frac{ \Gamma^{g,0}_\cusp}{\beta_0}\log \left(\frac{\alpha_s(\mu)}{\alpha_s(Q\sqrt{\tau})}\right)   \\
	&\ \ \times\exp\left[\frac{4\pi \Gamma^{g,0}_\cusp}{\beta_0^2 \alpha_s(Q\sqrt{\tau})} \left(\frac{\alpha_s(Q\sqrt{\tau})}{\alpha_s(\mu)} - 1 + 
	\log\frac{\alpha_s(\mu)}{\alpha_s(Q\sqrt{\tau})} \right) \right] .\nn
\end{align}
Therefore, as with the case of the soft function, our analytic solution of the subleading power mixing equation resums the logarithms at subleading power.

\section{Leading Logarithmic Resummation at Next-to-Leading Power}\label{sec:fact}

In this section we will apply the formalism for the resummation of subleading power jet and soft functions developed in the previous sections to resum the leading logarithms for thrust in pure glue $H\to gg$. This is a standard example used to study gluon jets.  We have chosen to restrict ourselves to the case of pure glue to demonstrate in the simplest setting the resummation of subleading power logarithms for a physical process and to highlight the role of the $\theta$-jet and $\theta$-soft operators and operator mixing. The inclusion of fermion operators and the extension to other processes is interesting, and will be considered in future work.

The complete structure of power corrections for dijet event shapes in SCET has been described in detail in the literature, where all relevant ingredients have been studied. In the effective theory, there are three sources of power corrections\footnote{The decomposition into these different classes of power corrections depends on the particular organization of the effective theory being used, but the final result does not.}
\begin{itemize}
\item Subleading power hard scattering operators \cite{Kolodrubetz:2016uim,Moult:2017rpl,Feige:2017zci,Chang:2017atu,Beneke:2017ztn}
\item Subleading power expansion of measurement operators and kinematics \cite{Freedman:2013vya,Feige:2017zci,Moult:2016fqy}
\item Subleading power Lagrangian insertions  \cite{Beneke:2002ni,Chay:2002vy,Manohar:2002fd,Pirjol:2002km,Beneke:2002ph,Bauer:2003mga,Moult:2019mog}
\end{itemize}
It was shown in \cite{Moult:2019mog} that there are no radiative contributions for pure glue $H\to gg$ at NLP at LL order. Therefore we need only consider the first two categories, namely hard scattering operators, and kinematic and measurement expansions, to derive the leading logarithms. We therefore write the cross section as
\begin{align}\label{eq:split}
\frac{1}{\sigma_0}\frac{\df\sigma^{(2)}_{\text{LL}} }{\df\tau} &= \frac{1}{\sigma_0}\frac{\df\sigma^{(2)}_{\kin,\text{LL}} }{\df\tau}+ \frac{1}{\sigma_0}\frac{\df\sigma^{(2)}_{\hard,\text{LL}} }{\df\tau} \,,
\end{align}
where we have put the subscript `LL` to emphasize that we will only give LL expressions for the factorization of the components, and will not include operators that first contribute at higher logarithmic order. In the next two sections we will explicitly work out the factorization and resummation for these two contributions. In both cases the resummation reduces to the mixing equation solved in \Sec{sec:solution}, allowing us to immediately derive the resummed result for thrust at subleading power. 

It is important to emphasize before continuing that the exact split between the terms in \Eq{eq:split} depends on the choice of momentum routing used to setup the factorization, although the final result for the factorization does not. For example, terms involving ultrasoft derivatives in $T$-products or hard scattering operators can in certain cases be eliminated from the hard term through a choice of momentum routing, and will then appear as kinematic corrections. However,  subleading power corrections from operators with additional ultrasoft fields are unambiguously in the hard component. We will define a convenient split in \Sec{sec:kin_corrections}.

\subsection{Kinematic and Observable Corrections}\label{sec:kin_corrections}

We begin by considering corrections from the expansion of the phase space (kinematics) and the thrust observable definition.
These were also considered in the fixed order calculations of \cite{Moult:2016fqy,Moult:2017jsg}, but here we will show how they can be treated to all orders as is required for factorization and resummation. In \cite{Feige:2017zci} it was shown through explicit calculation that the contributions from the thrust measurement function in our formalism do not contribute at LL order. We therefore only need to consider corrections to the phase space here.

\subsubsection{Factorization}\label{sec:kin_corrections_fact}

At subleading power, in addition to considering the expansion of the matrix elements which enter into the cross section, one must also consider power corrections arising from kinematic constraints on the phase space which can be neglected at leading power. To understand this issue we begin by writing the $N$ particle phase space
\begin{align}
\sigma = L_H \int \prod\limits_{i=1}^{N} \dbar^d p_i C(p_i) (2\pi)^4 \delta^4 \Bigl(q-\sum p_i\Bigr) |\cM|^2\,.
\end{align}
Here $q^2=Q^2$ is the momentum of the scattering, $\dbar^d p = d^d p/(2 \pi)^d$, $C(p)=2\pi \delta(p^2) \theta(p^0)$ is the on-shell particle constraint, and $L_H$ is the leptonic tensor. We now consider a final state consisting of $n$-collinear particles with total sector label mometum $\bar n \cdot k_n$, $\bar n$-collinear particles  with total sector label mometum $ n \cdot k_\bn$, and soft particles with total sector momentum $k_s$. Since $n\cdot k_s \sim \bar n \cdot k_s \sim \lambda^2$, at leading power, we can expand the momentum conserving delta function, and the incoming momentum $q$ fixes the large momentum of the collinear sector, namely
\begin{align}
\delta\Bigl(n\cdot q-\sum n\cdot p_i\Bigr)\, \delta\Bigl(\bn \cdot q-\sum \bn\cdot p_i\Bigr)
  =\delta(n\cdot q- n\cdot k_\bn)\, \delta(\bn \cdot q- \bn\cdot k_n)\,.
\end{align}
However, when working at subleading powers, we need to consider the power corrections to this formula, which we refer to as kinematic corrections. These can be organized in a number of different ways. Here we describe a way which seems particularly convenient for the process we are considering.

\begin{figure}
\begin{center}
\subfloat[]{\label{fig:routings_a}
  \includegraphics[width=0.37\textwidth]{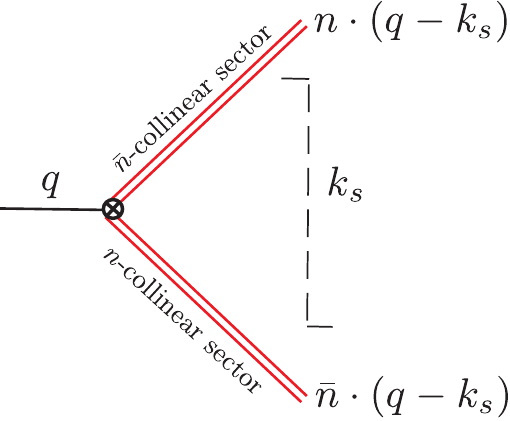}
  }
  \subfloat[]{\label{fig:routings_b}
  \includegraphics[width=0.35\textwidth]{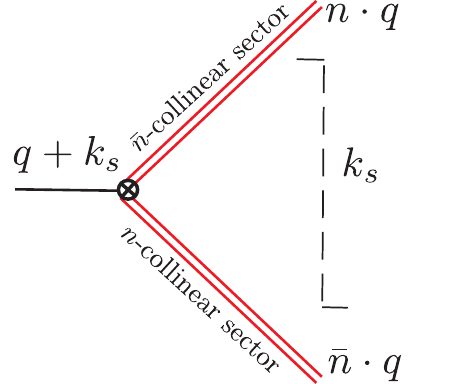}
  }
  \end{center}
  \caption{Two different routings for the soft momentum. In a) the additional soft momentum is routed into the collinear sectors. In b) the additional momentum is routed in through the hard scattering vertex, simplifying the large momentum routed into the collinear sectors.}
  \label{fig:routings}
\end{figure}

In SCET, exact momentum conservation for both label and residual components is implemented in all diagrams. 
Residual momenta must then be routed in the diagram, and unlike at leading power, their effects on the kinematics must be kept to the required power. This routing can be chosen arbitrarily, as long as it is done consistently for all contributions.\footnote{In particular, as mentioned above, this routing determines whether some contributions enter as kinematic or hard power corrections in the decomposition of \Eq{eq:split}.} As an example, consider the routing of the residual momentum from the soft sector. The most naive routing is shown in \Fig{fig:routings_a}. Here we imagine that the soft sector has a total momentum  $k_s$. This momentum must be extracted from the collinear sectors. The residual $n\cdot k_s\sim \lambda^2$ and $\bar n \cdot k_s \sim \lambda^2$ must be kept in the calculations of the collinear sector when working at $\cO(\lambda^2)$, complicating the calculations by requiring us to include $\partial_{us}$ acting on collinear lines.  Here we can still neglect the residual perp momentum of the soft sector, since this enters first as $k_\perp^2\sim \lambda^4$, which is beyond the order to which we work.

A more convenient routing is shown in \Fig{fig:routings_b}. Here, we instead route $q+k_s$ into the hard scattering vertex. The collinear sectors then have exactly $n\cdot q$ and $\bar n \cdot q$ as their large momentum contributions, and all kinematics in the final state is exact.  All kinematic corrections for this routing can be obtained by expanding the phase space factor in the leptonic tensor, which takes the form
\begin{align}\label{eq:lep_tensor_expand}
\frac{1}{(Q+ k_s)^4}=\frac{1}{Q^4}-2\frac{n\cdot k_s}{Q^5}-2\frac{\bar n\cdot k_s}{Q^5} +\cO(\tau^2)\,.
\end{align}
We therefore introduce the measurement functions
\begin{align}
n\cdot \hat k_s = \sum\limits_{i\in S} n\cdot k_s^i\,, \qquad  \bar n\cdot \hat k_s = \sum\limits_{i\in S} \bar n\cdot k_s^i\,,
\end{align}
where the sum is over all soft particles.
To LL accuracy we can make the replacement $n\cdot \hat k_s \to n\cdot \hat k_s \theta( \bn\cdot \hat k_s -n\cdot \hat k_s)$ and $\bn\cdot \hat k_s \to \bn\cdot \hat k_s \theta( n\cdot \hat k_s -\bn\cdot \hat k_s)$, since after multiplying the eikonal integrand $1/(l^+l^-)$ by $l^+$ (or $l^-$), the divergence responsible for the anomalous dimension comes only from the region of phase space where $l^-$ (or $l^+$) is unconstrained by the measurement. These kinematic corrections therefore combine to give the full thrust measurement function
\begin{align}
 n\cdot \hat k_s \theta( \bn\cdot \hat k_s -n\cdot \hat k_s)  + \bn\cdot \hat k_s \theta( n\cdot \hat k_s -\bn\cdot \hat k_s) = Q\hat \tau_s.
\end{align}
The $n\cdot k_n$  and $\bar n \cdot k_\bn$ residual momentum of each of the two collinear sectors can also be routed into the current in the exact same manner, leading to power correction given by $Q\hat \tau_n$ and  $Q\hat \tau_\bn$ respectively.

 We therefore find that the kinematic corrections arising from the phase space expansion give exactly the power suppressed jet and soft functions considered in \Sec{sec:renorm}, namely
\begin{align}
J^{(2)}_{g,\delta}(s, \mu)&=\frac{(2\pi)^3}{(N_c^2-1)}\langle 0 | \cB^{\mu a}_{n\perp}(0)\,\delta(Q+\bar \cP) \delta^2(\cP_\perp)\,  s~\delta\left(\frac{s}{Q}-\hat \Tau\right) \cB^{\mu a}_{n\perp}(0) |0 \rangle\,, \\
S^{(2)}_{g,\delta}(k, \mu)&=\frac{1}{(N_c^2-1)} \tr \langle 0 |  \cY^T_{\bar n}(0) \cY_n(0) k ~\delta(k-\hat \Tau) \cY_n^T(0) \cY_{\bar n}(0)|0\rangle\,. \nn
\end{align}
Indeed, this is one of the reasons why these particular subleading power jet and soft functions were used as an example in \Sec{sec:renorm}.

We can now write down an all orders factorization for the full contribution from kinematic corrections to the cross section at $\cO(\tau)$
\begin{align}\label{eq:fact_kin}
\frac{d\sigma_{\kin,\text{LL}}^{(2)}}{d\tau}&= n_\kin \int \frac{ds_n ds_\bn dk}{Q^2} \hat \delta_\tau    H^{(0)} (Q,\mu) J^{(2)}_{g,\delta} (s_n, \mu)  J_{g}^{(0)} (s_\bn, \mu)  S_{g}^{(0)}(k, \mu)   \\
   &+n_\kin \int \frac{ds_n ds_\bn dk}{Q^2} \hat \delta_\tau H^{(0)} (Q,\mu) J_{g}^{(0)} (s_n, \mu) J^{(2)}_{g,\delta}(s_\bn, \mu)   S_{g}^{(0)} (k, \mu) \nn  \\
& +n_\kin \int \frac{ds_n ds_\bn dk}{Q} \hat \delta_\tau H^{(0)}  (Q,\mu) J_{g}^{(0)} (s_n, \mu)  J_{g}^{(0)}(s_\bn, \mu)   S^{(2)}_{g,\delta}(k, \mu)  \,. \nn
\end{align}
The factorization for the kinematic corrections is therefore exactly the form considered in \Eq{eq:fact_NLP_multTau_rewrite}. 
We have explicitly put the subscript LL, to emphasize that beyond LL there would be additional contributions.
Here the integer constant
\begin{align}\label{eq:nkin_def}
n_\kin= -2\,,
\end{align}
is a normalization factor, effectively the number of times this contribution enters, which is obtained from \Eq{eq:lep_tensor_expand}. We have extracted it as a constant so as to be able to clearly track it, and distinguish it from other integer factors that will appear.

\subsubsection{Resummation}\label{sec:kin_corrections_renorm}

Since the kinematic contributions give exactly the illustrative example considered in \Sec{sec:renorm}, we can immediately perform the resummation of logarithms for this contribution using the solution to the mixing RG equation given in \Sec{sec:solution}. For concreteness, we can run both the soft and hard functions to the jet scale, $\mu_J =Q \sqrt{\tau}$ from their natural scales, $\mu_H=Q$ and $\mu_S=Q\tau$. At leading log order we can set $H^{(0)}(Q,Q)=1$ and $S^{(2)}_{g, \theta} (Q\tau,Q\tau) = \theta(\tau)$. We therefore have
\begin{align}\label{eq:kin_in_kernels}
\frac{1}{\sigma_0}\frac{\df\sigma^{(2)}_{\kin,\text{LL}} }{\df\tau}&=-2 \,U_H(Q, Q\sqrt{\tau})U^S_{g,\delta\theta}(Q\tau,Q \sqrt{\tau})\, \theta(\tau)\,, 
\end{align}
Here the hard evolution kernel is that of the leading power hard function.
\begin{align}\label{eq:LPHkernel}
U_H(Q,Q\sqrt{\tau}) = \exp\left\{-\frac{4\pi \Gamma^{g,0}_\cusp}{\beta_0^2\alpha_s(Q)} \left[ \frac{\alpha_s(Q)}{\alpha_s(Q\sqrt{\tau})} - 1 + \log \left( \frac{\alpha_s(Q\sqrt{\tau})}{\alpha_s(Q)}\right) \right] \right\} \,.
\end{align}
where $ \Gamma^{g,0}_\cusp =4C_A$ is the one-loop gluon cusp anomalous dimension.
The resummed soft function is given by the combination
\begin{align}
	S^{(2)}_{g, \delta}(Q\tau, \mu=Q \sqrt{\tau}) 
=  U^S_{g,\delta\theta}(Q\tau,Q \sqrt{\tau})\, 
  S^{(2)}_{g, \theta} (Q\tau,\mu_0=Q\tau)\,,
\end{align}
and by taking the result of \Eq{eq:SresummedQt} with $\mu = Q\sqrt{\tau}$, we have that the evolution kernel for the soft function at LL reads
\begin{align}\label{eq:SkinLLrunning}
	U^{S\,\LL}_{g,\delta\theta}(Q\tau,Q\sqrt{\tau}) &= - \frac{2 \Gamma^{g,0}_\cusp}{\beta_0}\log \left(\frac{\alpha_s(Q\sqrt{\tau})}{\alpha_s(Q\tau)}\right) \\
	&\ \ \times \exp\left\{-\frac{4\pi \Gamma^{g,0}_\cusp}{\beta_0^2 \alpha_s(Q\tau)} \left[\frac{\alpha_s(Q\tau)}{\alpha_s(Q\sqrt{\tau})} - 1 + \log  \left(\frac{\alpha_s(Q\sqrt{\tau})}{\alpha_s(Q\tau)}\right)\right] \right\} \nn\,.
\end{align}
Plugging these expressions for the evolution kernels into \Eq{eq:kin_in_kernels}, we find that the resummed result for the kinematic contributions is given by
\begin{align}\label{eq:SkinLLrunning_b}
	\frac{1}{\sigma_0}\frac{\df\sigma^{(2)}_{\kin,\text{LL}} }{\df\tau}&= \theta(\tau) \frac{4 \Gamma^{g,0}_\cusp}{\beta_0}\log \left(\frac{\alpha_s(Q\sqrt{\tau})}{\alpha_s(Q\tau)}\right) \exp\biggl\{-\frac{4\pi \Gamma^{g,0}_\cusp}{\beta_0^2 } \biggl[\frac{2}{\alpha_s(Q\sqrt{\tau})} - \frac{1}{\alpha_s(Q\tau)} - \frac{1}{\alpha_s(Q)} \nn\\
	&\quad +  \frac{1}{\alpha_s(Q\tau)}\log  \left(\frac{\alpha_s(Q\sqrt{\tau})}{\alpha_s(Q\tau)}\right) + \frac{1}{\alpha_s(Q)}\log  \left(\frac{\alpha_s(Q\sqrt{\tau})}{\alpha_s(Q)}\right)\biggr] \biggr\}\,.
\end{align}
Simplifying to the case of a fixed coupling and plugging in $\Gamma^{g,0}_\cusp=4C_A$, the kinematic contribution at leading log reads
\begin{align}
\frac{1}{\sigma_0}\frac{\df\sigma^{(2)}_{\kin,\text{LL}} }{\df\tau}&=\left(  \frac{\alpha_s}{4\pi} \right)16 C_A \theta(\tau)\log(\tau) e^{- \frac{\alpha_s}{4\pi} \Gamma^{g,0}_{\text{cusp}} \log^2(\tau)} \,. 
\end{align}
This is a remarkably simple result, involving double logarithmic asymptotics governed by the cusp anomalous dimension. However, this is not surprising since these corrections arise from a multiplication of the leading power result by $\tau$.

\subsection{Hard Scattering Operators}\label{sec:hard_operators}

The second class of contributions that are required for the LL description at NLP arise from corrections to the scattering amplitudes themselves, which in this case are described by subleading power hard scattering operators in the EFT.  A complete basis of hard scattering operators at $\cO(\lambda^2)$ for $H\to gg$ was derived in \cite{Moult:2017rpl}. 

At subleading powers, it becomes important to work in terms of gauge invariant fields, even at the ultrasoft scale.  Leading power interactions between soft and collinear particles in the effective theory can be decoupled to all orders using the BPS field redefinition \cite{Bauer:2002nz}, which for the gluon operator reads
\be \label{eq:BPSfieldredefinition}
\cB^{a\mu}_{n\perp}\to \cY_n^{ab} \cB^{b\mu}_{n\perp}\,.
\ee
This factorizes the Hilbert space into separate soft and collinear sectors. After performing the BPS field redefinition, operators in the effective theory can be written in terms of gauge invariant soft and collinear gluon fields
\begin{align} \label{eq:softgluondef}
g \cB^{a\mu}_{us(i)}&= \left [   \frac{1}{in_i\cdot \partial_{us}} n_{i\nu} i G_{us}^{b\nu \mu} \cY^{ba}_{n_i}  \right] \,, \qquad g\cB_{n_i\perp}^{A\mu} =\left [ \frac{1}{\bar \cP}    \bar n_{i\nu} i G_{n_i}^{B\nu \mu \perp} \cW^{BA}_{n_i}         \right]\,,
\end{align}
where $\cY$ and $\cW$ are adjoint soft and collinear Wilson lines (see \Eq{eq:Wilson_def}). Due to the presence of the Wilson lines, these gauge invariant fields have Feynman rules at every order in $\alpha_s$. An identical construction exists for collinear and soft fermions, although they will not be needed here since we focus on pure Yang-Mills theory.

The subleading power operators that contribute to the LL cross section involve either an insertion of the $\cB_{n\perp}$, or $\cB_{us}$ operators. The relevant operators, along with their tree level matching coefficients which are required for LL resummation, are given in \Tab{tab:ops}.  The leading power operator is also given for convenience. An important simplification which occurs for the soft operators is that their Wilson coefficients are fixed by reparametrization invariance (RPI) \cite{Larkoski:2014bxa}. In particular, we have the all orders relation
\begin{align} \label{eq:usRPIrelation}
C^{(2)}_{\cB \bn(us)}&=-\frac{\partial C^{(0)} }{\partial \omega_1} 
\,, 
\end{align}
and similarly for $n\leftrightarrow \bar n$. As we will see, this will provide a significant simplification, since it fixes the anomalous dimensions of these soft operators. This relationship can be viewed as a manifestation of the Low-Burnett-Kroll theorem \cite{Low:1958sn,Burnett:1967km}, where the connection with our SCET based approach has been explained in detail in \cite{Larkoski:2014bxa}.

{
\renewcommand{\arraystretch}{1.4}
\begin{table}[t!]
\begin{center}
\scalebox{0.842}{
\begin{tabular}{| l | c | c |c |c|c| r| }
  \hline                       
   Operator & Tree Level Matching Coefficient \\
  \hline
    $\cO_\cB^{(0)}=C^{(0)} \delta^{ab} \cB_{\perp \bar n, \omega_2}^a \cdot \cB_{\perp \bar n, \omega_1}^b H$& $C^{(0)}=-2\omega_1 \omega_2\,.$ \\
  \hline
   $\cO^{(2)}_{\cP \cB1}=C^{(2)}_{\cP \cB1}i f^{abc} \cB^a_{n\perp,\omega_1}\cdot \left[  \cP_\perp \cB^b_{\bar n \perp,\omega_2}\cdot  \right] \cB_{\bar n \perp,\omega_3}^c    H$ &  $C^{(2)}_{\cP \cB1}=-\left( \frac{1}{2}\right)4g \left(  2+\frac{\omega_3}{\omega_2}+ \frac{\omega_2}{\omega_3}  \right)$ \\
   \hline
   $\cO^{(2)}_{\cP \cB2}=C^{(2)}_{\cP \cB2} if^{abc}\left[ \cP_\perp \cdot \cB_{\bar n \perp,\omega_3}^a \right] \cB^b_{n\perp,\omega_1} \cdot \cB_{\perp \bar n, \omega_2}^c    H$ & $C^{(2)}_{\cP \cB2}=4g\left( 2+\frac{\omega_3}{\omega_2}  + \frac{\omega_2}{\omega_3}\right)$\\
  \hline  
  $\cO^{(2)}_{\cB(us(n))}=C^{(2)}_{\cB \bn(us)} \left(i  f^{abd}\, \big({\cal Y}_n^T {\cal Y}_{\bar n}\big)^{dc}\right)  \left (  \cB^a_{n\perp, \omega_1} \cdot \cB^b_{\bar n \perp, \omega_2} \bar n \cdot g\cB^c_{us(n)} \right)$ & $C^{(2)}_{\cB \bn(us)}=-2 \omega_2$ \\
  \hline
  $\cO^{(2)}_{\cB(us(\bar n))}= C^{(2)}_{\cB n(us)} \left(i  f^{abd}\, \big({\cal Y}_{\bar n}^T {\cal Y}_{n}\big)^{dc}\right)  \left ( \cB^a_{n\perp, \omega_1} \cdot \cB^b_{\bar n \perp, \omega_2} n\cdot g\cB^c_{us(\bar n)} \right)$ & $C^{(2)}_{\cB n(us)}=-2\omega_1$ \\
  \hline
\end{tabular}}
\end{center}
\caption{
Hard scattering operators that contribute to the LL cross section to $\cO(\lambda^2)$, along with their tree level matching coefficients. These operators and matching coefficients were derived in \cite{Moult:2017rpl}.
}
\label{tab:ops}
\end{table}
}

The operators which contribute to the fixed order leading logarithms were identified in the calculation of  \cite{Moult:2017jsg} as those which contribute a logarithm at the lowest order in perturbation theory. The leading logarithms to all orders are then obtained by the renormalization of these contributions, which dresses them with an all orders resummation of double logarithms. To prove that this is indeed the case, we can assume that there exists a jet or soft function that first contributes at some higher order, for concreteness $\alpha_s^2$,  and that this contribution is leading logarithmic, and hence contributes as $\alpha_s^2 \log^3(\tau)$. With our understanding of the renormalization of subleading jet and soft functions, we know that this implies that this function must be renormalized by a subleading power $\theta$-function type operator, since it can't be a self renormalization. Taking $\mu d/d\mu$, the anomalous dimension of such a LL mixing contribution would have to be of the form $\gamma \sim \log^2(\mu/\mu_0)$. However, it is know that anomalous dimensions in SCET can be at most linear in logarithms, which is required by RG consistency. This argument was first presented in \cite{Manohar:2003vb} in the context of leading power RG consistency. Since this argument relies only on the additive properties of the logarithm, it applies also here. This implies that the operators appearing in \Tab{tab:ops} are sufficient to derive the LL resummation.

\subsubsection{Factorization}\label{sec:hard_operators_factorization}

With an understanding of the operators that contribute, it is now straightforward to write down a factorization for their contributions, which is sufficient for the LL resummation. Detailed accounts of the factorization of matrix elements at subleading power have been given in \cite{Lee:2004ja,Beneke:2004in,Hill:2004if,Freedman:2013vya,Moult:2019mog}. Since the focus of this paper is on the LL resummation through the mixing with the $\theta$-jet and $\theta$-soft operators, here we simply present the final result for the factorization. Since there are only a small number of operators that appear due to our restriction to a pure glue final state we find a simple LL factorization formula
\begin{align}\label{eq:fact_hard}
\frac{1}{\sigma_0}\frac{d\sigma_{\hard,\text{LL}}^{(2)}}{d\tau}&=n_\hard \int  \frac{ds_n ds_\bn dk}{Q}  \hat \delta_\tau H_{n \cdot \cB}(Q,\mu)     S_{\bn\cB_{us}}^{(2)}(k, \mu)J^{(0)}_{g}(s_n, \mu)\ J^{(0)}_{g}(s_\bn,\mu) \\
&+  n_\hard \int \frac{ds_n ds_\bn dk}{Q^2} \hat \delta_\tau \int d\omega~ H_{\cB\cP}(\omega, Q, \mu) S_{g}^{(0)}(k, \mu) J_{\cB\cP}^{(2)}(s_\bn,\omega, \mu) J_{g}^{(0)}(s_n, \mu)\,. \nn
\end{align}
Here 
\begin{align}\label{eq:nhardvalue}
n_\hard=2\,,
\end{align}
is a combinatorial factor from the equality of $S_{\bn \cB_{us}}^{(2)}$ and $S_{n \cB_{us}}^{(2)}$ in the first line, and from correcting both jet functions and taking the symmetric combination in the second.
This factorization involves a power suppressed soft function
\begin{align}
S_{\bn \cB_{us}}^{(2)}(k, \mu)=&\frac{if^{abd}}{N_c^2-1}\tr \langle 0 |  (\cY_n^T(0) \cY_\bn(0))^{dc} \bar n \cdot g\cB^c_{ us (n)}(0) \delta(k-\hat \Tau) (\cY_n(0) \cY^T_{\bar n}(0))^{ab} |0 \rangle \,,
\end{align}
which arises from the insertion of the $\cB_{us}$ field into the standard leading power soft function. Here we have absorbed the $g$ from the matching coefficient into the soft function. As with the previous subleading power soft functions we have defined in \Eqs{eq:tau_funcs}{eq:theta_soft_first}, this subleading power soft function has mass dimension zero. This factorization also involves a  subleading power jet function
\begin{align}
 \cJ^{(2)}_{\cB\cP}(s, \omega,\mu)& =\\
 &\hspace{-1cm}\frac{(2\pi)^3}{(N_c^2-1)}\frac{Q^2}{\omega (Q-\omega)}\langle 0| [\cB_{\perp \bar n,\omega}(0) [g\cB_{\perp \bar n}(0) \cdot \cP_\perp^\dagger] \delta(Q+\bar \cP) \delta^2(\cP_\perp)\, \delta\left(\frac{s}{Q}-\hat \Tau\right)  \cB_{\perp \bar n}(0) |0\rangle\,,\nn
\end{align}
which arises from the hard scattering operators involving an additional $\cB_{\perp}$ field, and $\cP_\perp$ operator. We have again absorbed the $g$ from the matching coefficient into the definition of the jet function, and as with the subleading power jet functions of \Eqs{eq:tau_funcs}{eq:theta_op} we have defined this jet function to have mass dimension 0.  This jet function involves a convolution in an additional label variable, which is the label momentum of one of the $\cB_\perp$ fields. However, at LL this does not play a role in its renormalization.

\subsubsection{Resummation}\label{sec:hard_operators_renormalization}

Using the factorized expression for the hard scattering operators, we can resum their contribution to the cross section to LL accuracy. To simplify the LL analysis as much as possible, we can exploit consistency relations in the RG equations. As mentioned in \Sec{sec:consistency}, since the subleading power jet and soft functions start as $\cO(\alpha_s)$, we can always choose to eliminate one of them. In the present case, it is convenient to choose to run to the jet scale, where 
\begin{align}
J_{\cB\cP}^{(2)}(s,\omega,\mu)&= 0+\cO(\alpha_s)\,.
\end{align}
With this choice, we do not need to consider the power suppressed jet functions. 

We do, however, have to consider the renormalization of the subleading power soft functions, and the hard function $H_{\bn \cdot \cB}$. However, as described in \Sec{sec:hard_operators}, the anomalous dimension of this hard function is fixed by RPI due to the relation of \Eq{eq:usRPIrelation}. This can be seen by differentiating the RG equation for the leading power Wilson coefficient, whose all orders structure is
\begin{align}
\mu \frac{d}{d\mu}C^{(0)} (\omega_1, \omega_2, \mu)=\gamma_C (\omega_1, \omega_2, \mu) C^{(0)} (\omega_1, \omega_2, \mu)\,.
\end{align}
Taking the derivative with respect to $\omega_1$, and switching the order of differentiation, we find 
\begin{align}
\mu \frac{d}{d\mu} \left[  \frac{\partial}{\partial\omega_1} C^{(0)} (\omega_1, \omega_2, \mu)   \right]& = \frac{\partial}{\partial\omega_1} [\gamma_C (\omega_1, \omega_2, \mu)]  C^{(0)} (\omega_1, \omega_2, \mu)  \\
&+ \gamma_C(\omega_1, \omega_2, \mu) \frac{\partial}{\partial\omega_1}   C^{(0)}  (\omega_1, \omega_2, \mu)\,.\nn
\end{align}
The all orders form of the anomalous dimension for the leading power matching coefficient is given by
\begin{align}\label{eq:hard_RG_LP}
\gamma_C (\omega_1,\omega_2,\mu) = \Gamma^g_\cusp[\alpha_s(\mu)] \log \left( \frac{-\omega_1 \omega_2}{\mu^2}  \right) +\gamma_C[\alpha_s(\mu)]\,,
\end{align}
where the second term $\gamma_C[\alpha_s(\mu)]$ is the non-cusp anomalous dimension, which contains no logarithms, and drives the single logarithmic evolution.
The leading double logarithmic evolution is governed by the cusp component. The differentiation in the first component removes the double log component, and therefore we have that to LL accuracy
\begin{align}
\mu \frac{d}{d\mu} \left[  \frac{\partial}{\partial\omega_1} C^{(0)} (\omega_1, \omega_2, \mu)  \right] = \gamma_C(\omega_1, \omega_2, \mu) \left[ \frac{\partial}{\partial\omega_1}  C^{(0)}  (\omega_1, \omega_2, \mu) \right]\,.
\end{align}
This shows that the  LL RG evolution for the subleading power hard scattering operators involving a $\cB_{{us}}$ is identical to that for the leading power hard function, and in particular, is driven by the cusp anomalous dimension.

Finally, the self mixing anomalous dimension of the subleading power soft function is also fixed by RG consistency. In particular, the jet functions appearing in the factorization of \Eq{eq:fact_hard} are the leading power jet functions, and their anomalous dimensions are given in \Eq{eq:LP_anom_dim}. Combining this with the known anomalous dimension for the hard function, it implies by RG consistency relations of \Sec{sec:consistency} that the self mixing anomalous dimension of the subleading power soft function is equal to that of the leading power soft function to LL.

We therefore only need to compute the mixing anomalous dimensions into the $\theta$ function operators for the soft functions involving the $\cB_{{us}}$ operators.  Computing the one loop matrix element of the power suppressed soft function, we find
\begin{align}
\left. S_{\bn \cB_{us}}^{(2)}(k, \mu)\right|_{\cO(\alpha_s)}&= \fd{2cm}{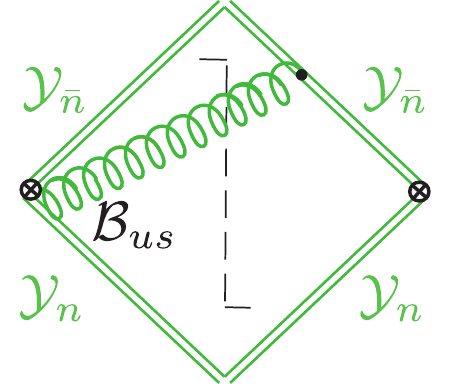}+\fd{2cm}{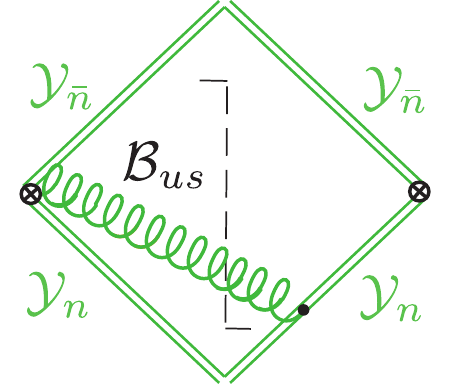} \\
&=g^2  \left(  \frac{\mu^2 e^{\gamma_E}}{4\pi} \right)^\epsilon  C_A \int \frac{d^dl}{(2\pi)^d} \left( \frac{2}{l^+}+ \frac{2}{l^-} \right )2\pi\delta(l^2) \theta(l^0) k \delta(k-Q\hat \tau) \nn\\
&=4C_A\frac{\alpha_s(\mu)}{4\pi}\theta(k) \left(   \frac{1}{\epsilon} +  \log \left(  \frac{\mu^2}{k^2}\right)  +\cO(\epsilon)\right)\,. \nn
\end{align}
As with the illustrative example of \Eq{sec:renorm}, we see that this soft function mixes with a $\theta$-function operator. The RG consistency  relations of \Sec{sec:consistency} imply that the all orders structure of the function being mixed into is that of the adjoint soft function $\theta$-function operator of \Eq{eq:theta_soft_first}. We note that this is a highly non-trivial statement, which would be difficult to prove in perturbation theory, but is dictated by the RG consistency equations of the EFT. We therefore find a $2\times 2$ mixing structure
\begin{align}\label{eq:2by2mix_S_mod}
\mu \frac{d}{d\mu}\left(\begin{array}{c} S_{\bn\cB_{us}}(k,\mu) \\ S_{g,\theta}(k,\mu) \end{array} \right) &= \int dk' \left( \begin{array}{cc} \gamma^S_{\bn \cdot \cB_{us}}(k-k',\mu) & \gamma_{\bn\cdot \cB_{us} \theta}\, \delta(k-k') \\  0 &  \gamma^S_{g,\theta \theta}(k-k',\mu)   \end{array} \right) \left(\begin{array}{c} S_{\bn\cB_{us}}(k',\mu) \\ S_{g, \theta}(k',\mu) \end{array} \right) \,,
\end{align}
where to LL accuracy,
\begin{align}\label{eq:anom_dim_Bus}
\gamma^S_{\bn \cdot \cB_{us}}(k,\mu)&= 4 \Gamma_{\text{cusp}}^{g0} \frac{\alpha_s(\mu)}{4\pi} \, \frac{1}{\mu} \biggl[  \frac{\mu\,\theta(k)}{k}  \biggr]_+\,, \\
\gamma_{\bn\cdot\cB_{us} \theta}&= 8C_A \frac{\alpha_s(\mu)}{4\pi}  \,.\nn
\end{align}
This therefore determines all the anomalous dimensions that are required for LL resummation at NLP. Since the RG equation takes exactly the form already solved in \Sec{sec:solution}, we can immediately use those results to perform the resummation.

Just as for the kinematic contribution, here we run all the functions to the jet scale, $\mu_J^2 =Q^2 \tau$. At their natural scales, $\mu_H=Q$ and $\mu_S=Q\tau$, the hard and the soft function are respectively\footnote{$H_{\bar n \cdot \cB}$ is related to the Wilson coefficient $C^{(2)}_{\cB \bn(us)}$ of the hard scattering operator. From Table \ref{tab:ops} we see that at LP we have $|C^{(0)}(Q,Q)|^2 = 4 Q^4$, and these factors are contained in the normalization factor $\sigma_0$. At subleading power this factor is coming from the interference of $O^{(2)}_{\cB \bn(us)}$ with $O^{(0)}$, which gives $C^{(2)}_{\cB \bn(us)}(Q,Q)C^{(0)}(Q,Q) = 4 Q^3$. In \Eq{eq:fact_hard} one can see the extra ${1}/{Q}$ in the prefactor of the factorization theorem which is precisely the ratio of the tree level subleading Wilson coefficient by the LP one. Thus our $H_{\bar n \cdot \cB}(Q,Q)$ is normalized so that it is dimensionless and equal to $1$ at tree level.} $H_{\bar n \cdot \cB}(Q,Q) = 1$ and  $S^{(2)}_{g, \theta} (Q\tau,Q\tau) = \theta(\tau)$. Using $n_\hard = 2$ from \Eq{eq:nhardvalue}, the hard scattering operator contribution is
\begin{align}
\frac{1}{\sigma_0}\frac{\df\sigma^{(2)}_{\hard,\text{LL}} }{\df\tau}
  &=2\, U_{H_{\bar n \cdot \cB}}(Q, Q \sqrt{\tau}) \,
    U^S_{\bn  B_{us}}(Q\tau,Q \sqrt{\tau}) \, \theta(\tau) \,.
\end{align}
As was shown above, the hard evolution kernel $U_{H_{\bar n \cdot \cB}}(Q, Q \sqrt{\tau})$ is identical to that for the leading power operator, which is quoted in \Eq{eq:LPHkernel}. The soft function takes an identical form to that given in \Eq{eq:Fmomspacetext}, but with $k=\mu_0= Q\tau$ and the anomalous dimensions from \Eq{eq:anom_dim_Bus}. Hence, we get
\begin{align}
S^{(2)\LL}_{g,\delta}(Q\tau,Q\sqrt{\tau}) &= -\theta(\tau) \frac{8 C_A}{\beta_0}\log (r) \,\exp\left[-\frac{4\pi \Gamma^{g,0}_\cusp}{\beta_0^2 \alpha_s(Q\tau)} \left(\frac{1}{r} - 1 + \log(r)\right) \right] \,, 
\end{align} 
where here we have
\begin{align}
 r = \frac{\alpha_s(Q\sqrt{\tau})}{\alpha_s(Q\tau)}\,.
\end{align}
Combining these pieces together, we have 
\begin{align}
	\frac{1}{\sigma_0}\frac{\df\sigma^{(2)}_{\hard,\text{LL}} }{\df\tau}&= -\theta(\tau) \frac{2 \Gamma^{g,0}_\cusp}{\beta_0}\log \left(\frac{\alpha_s(Q\sqrt{\tau})}{\alpha_s(Q\tau)}\right) \exp\biggl\{-\frac{4\pi \Gamma^{g,0}_\cusp}{\beta_0^2 } \biggl[\frac{2}{\alpha_s(Q\sqrt{\tau})} - \frac{1}{\alpha_s(Q\tau)} - \frac{1}{\alpha_s(Q)} \nn\\
	&\quad +  \frac{1}{\alpha_s(Q\tau)}\log  \left(\frac{\alpha_s(Q\sqrt{\tau})}{\alpha_s(Q\tau)}\right) + \frac{1}{\alpha_s(Q)}\log  \left(\frac{\alpha_s(Q\sqrt{\tau})}{\alpha_s(Q)}\right)\biggr] \biggr\} \,.
\end{align}
As with the kinematic contribution to the cross section, we find that the contribution from hard scattering operators resums at LL accuracy into a Sudakov exponential governed by the cusp anomalous dimension.

It is important to emphasize that the simplicity of this result is largely due to the restriction to LL. At LL accuracy the anomalous dimensions do not involve additional convolution variables in the subleading power jet and soft functions, and are purely multiplicative in these variables. This significantly simplifies the structure, with the primary ingredient to achieve renormalization and resummation being the mixing with the $\theta$-jet and $\theta$-soft functions. Beyond LL, the $\theta$-jet and $\theta$-soft will continue to play an important role, but the convolution structure will become more complicated. 

\subsection{Resummed Result for Thrust in $H\to gg$ at Next-to-Leading Power}\label{sec:resum}

\begin{figure}
\begin{center}
  \vspace{-0.2cm}
  \label{fig:sudakovs_a}
  \raisebox{-0.5cm}{a)}\hspace{-1cm}
  \includegraphics[width=0.49\textwidth]{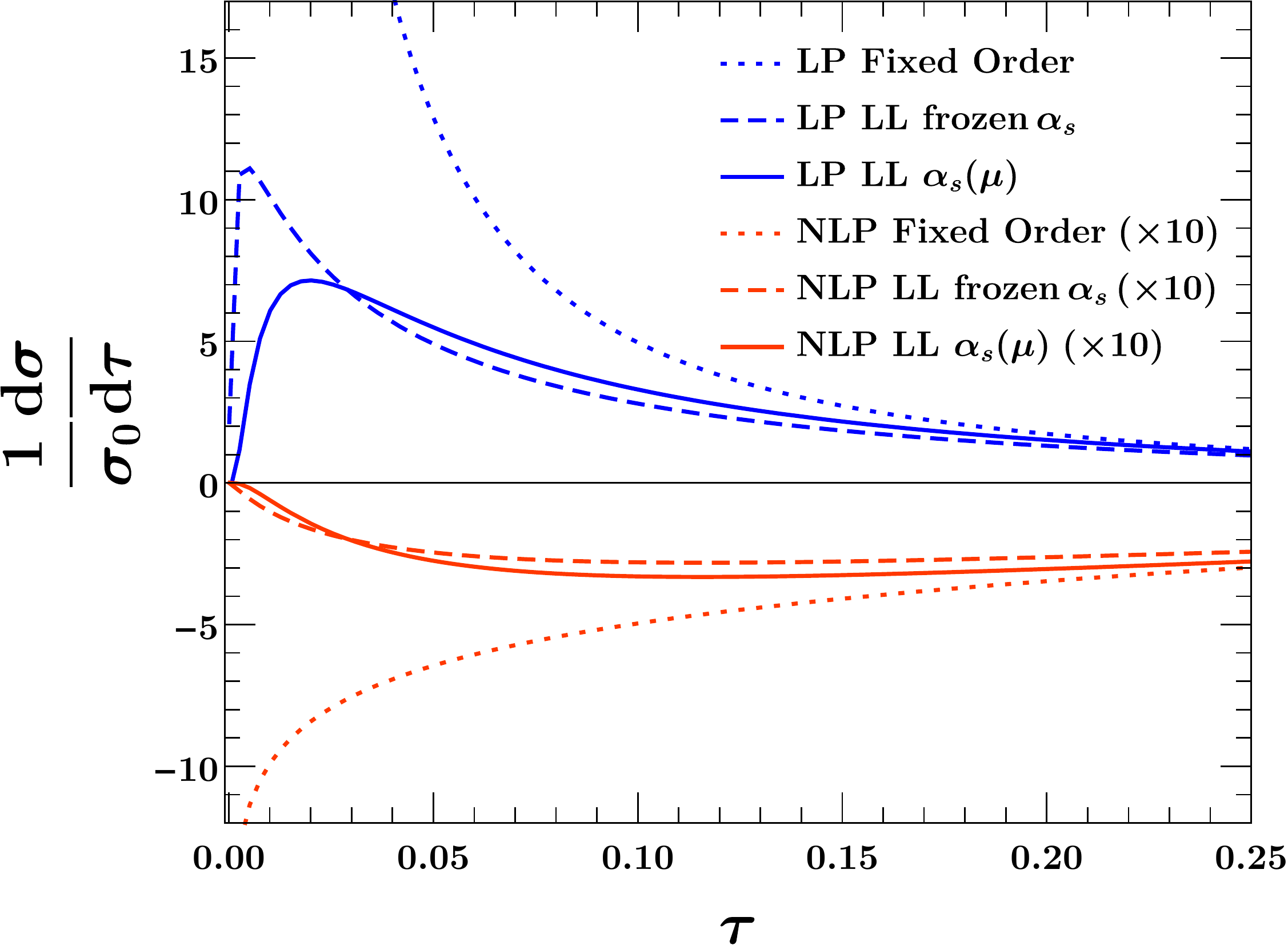}
  \hspace{1cm}\raisebox{-0.5cm}{b)}\hspace{-1.2cm} 
  \includegraphics[width=0.505\textwidth]{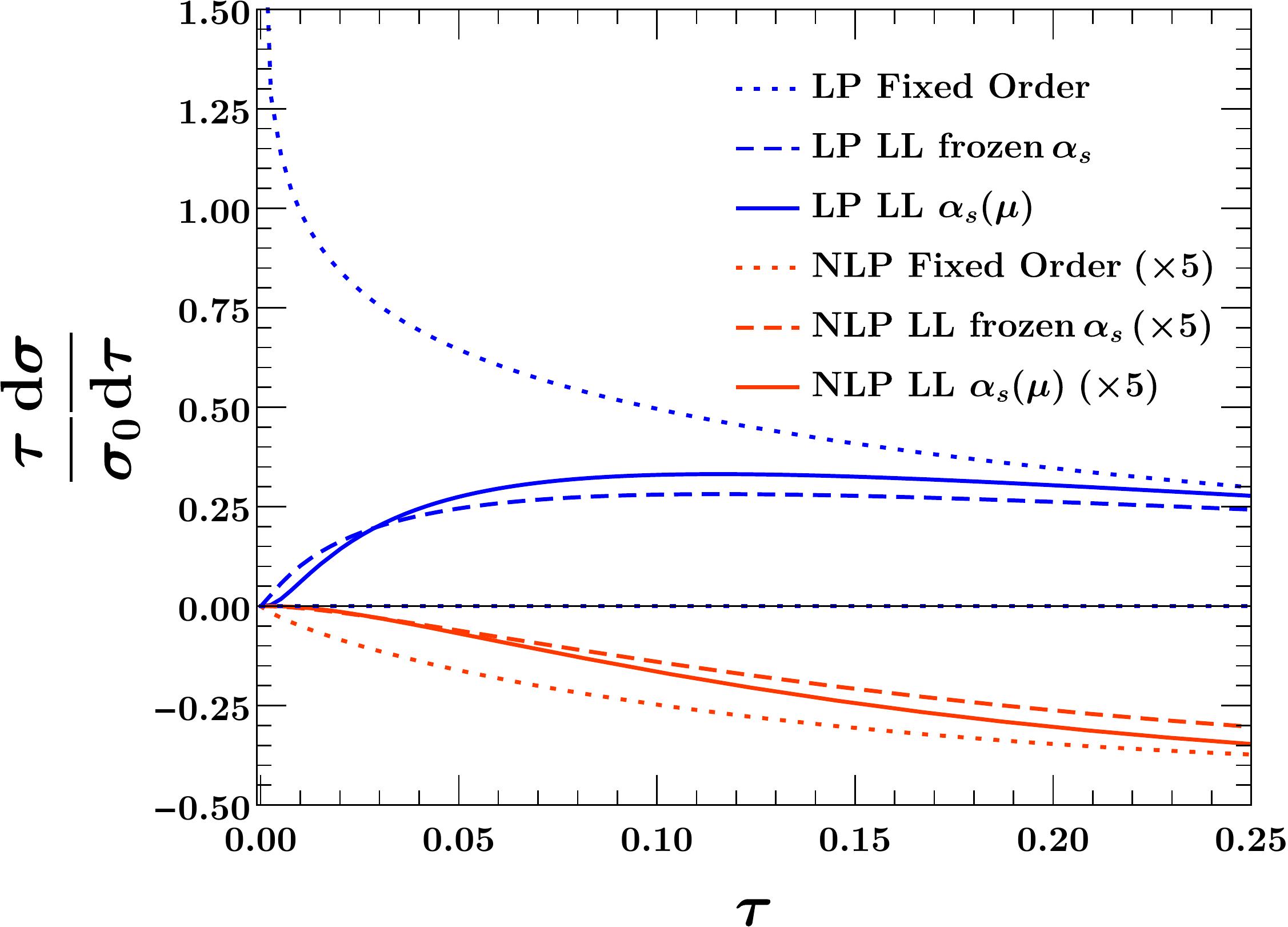}
  \end{center}
\vspace{-0.2cm}
  \caption{Plots of the LP and NLP fixed order and resummed predictions for thrust in pure glue $H\to gg$, with and without running coupling. In a) we show $d\sigma/d\tau$ and in b) we show $\tau d\sigma/ d\tau$. Resummation at LP cures a $1/\tau$ divergence, while resummation at NLP overturns a much weaker logarithmic divergence, leading to a broader shoulder.}
  \label{fig:sudakovs}
\end{figure}

Having resummed the two different contributions to the cross section in \Eq{eq:split}, we can now give a resummed result for thrust in pure glue $H\to gg$.
 Adding together the different contributions, each of which is dressed by the same Sudakov exponential, we find
\begin{align}\label{eq:resum_nlp}
\frac{1}{\sigma_0}\frac{\df\sigma^{(2)}_{\text{LL}} }{\df\tau} &= \frac{1}{\sigma_0}\frac{\df\sigma^{(2)}_{\kin,\text{LL}} }{\df\tau}+ \frac{1}{\sigma_0}\frac{\df\sigma^{(2)}_{\hard,\text{LL}} }{\df\tau} \nn\\
&= \theta(\tau)\frac{8C_A}{\beta_0}\log \left(\frac{\alpha_s(Q\sqrt{\tau})}{\alpha_s(Q\tau)}\right) \exp\biggl\{-\frac{4\pi \Gamma^{g,0}_\cusp}{\beta_0^2 } \biggl[\frac{2}{\alpha_s(Q\sqrt{\tau})} - \frac{1}{\alpha_s(Q\tau)} - \frac{1}{\alpha_s(Q)} \nn\\
	&\quad +  \frac{1}{\alpha_s(Q\tau)}\log  \left(\frac{\alpha_s(Q\sqrt{\tau})}{\alpha_s(Q\tau)}\right) + \frac{1}{\alpha_s(Q)}\log  \left(\frac{\alpha_s(Q\sqrt{\tau})}{\alpha_s(Q)}\right)\biggr] \biggr\}\,.
\end{align}
With a fixed coupling, \Eq{eq:resum_nlp} simplifies to
\begin{align}\label{eq:resum_nlp_fixed}
\frac{1}{\sigma_0}\frac{\df\sigma^{(2)}_{\text{LL}} }{\df\tau}\Big|_{\alpha_s(\mu)=\alpha_s} &= \left(  \frac{\alpha_s}{4\pi} \right) 8C_A \theta(\tau) \log(\tau) e^{-4C_A \frac{\alpha_s}{4\pi} \log^2(\tau)}   \,. 
\end{align}
This shows the exponentiation of the subleading power logarithms into a Sudakov form factor governed by the cusp anomalous dimension, and is one of the main results of this paper. We note that this result is simply $-\tau$ multiplying the LP result with LL resummation. This simplicity is in part related to the fact that we have chosen a simple event shape example, and is not expected to hold in general at LL, nor beyond LL.
In \Sec{sec:split} we will check this result to $\cO(\alpha_s^3)$ by expanding known results for the amplitudes \cite{Garland:2001tf,Garland:2002ak,Gehrmann:2011aa}, and find complete agreement.

This resummation tames the (integrable) singularity in the subleading power cross section as $\tau \to 0$. A plot of the LL NLP resummed cross section is shown in \Fig{fig:sudakovs}, along with the NLP fixed order results, and the LP results. Results with and without running coupling are shown. We use $\alpha_s(m_Z)=0.118$ for the running coupling $\alpha_s(\mu)$, and when we freeze the coupling, we use $\alpha_s=\alpha_s(m_H)=0.113$.  The NLP results are multiplied by a factor of 10 in \Fig{fig:sudakovs} a) and a factor of 5 in \Fig{fig:sudakovs} b) to make them visible. Due to the fact that the NLP result is not enhanced by a factor of $1/\tau$ it leads to a much broader result, peaked at large values of $\tau$. This has interesting consequences for the effect of the running coupling. In particular, at subleading power the running coupling has a much smaller effect, since the distribution is more suppressed at smaller values of $\tau$.  At higher powers, resummation is not required for the cross section to go to zero as $\tau \to 0$, since the corrections behave as $\tau^n \log^m(\tau)$, with $n>0$. Nevertheless,  RG equations are still useful for predicting higher order terms in the perturbative expansion.

\section{Subleading Power Collinear Limit and Fixed Order Check}\label{sec:split}

In this section we check our resummed result for thrust to $\cO(\alpha_s^3)$ by explicitly calculating the power corrections to this order. This is achieved by exploiting a relation between the LL result and the subleading power collinear limit of the involved amplitudes. We also discuss flipping around this logic, and using the resummed results to constrain corrections in the collinear limit at $n$th-loop order. In particular, for $H\to ggg$ we will show that the same loop corrections dress terms that appear at leading and next-to-leading order in the power expansion. 

The $N$-loop fixed order result at NLP can be written as \cite{Moult:2016fqy,Moult:2017jsg}
\begin{align}\label{eq:constraint_setup}
\frac{1}{\sigma_0}\frac{\df\sigma^{(2,N)}}{\df\tau}
  = & \sum_{\kappa}\sum_{i=0}^{2N-1} \frac{c_{\kappa,i}}{\epsilon^i} \left( \frac{\mu^{2N}}{Q^{2N} \tau^{m(\kappa)}}   \right)^\epsilon
 + \dots
\,,\end{align}
where the dots involve terms that are first relevant beyond LL order. Our superscript $(j,N)$ notation denotes the subleading power at order $j$ and loop order $N$. Here the sum over $\kappa$ is over different possible combinations of soft, collinear, or hard particles entering the $N$-loop result, and the power $m(\kappa)$ appearing in \Eq{eq:constraint_setup} depends on this combination. For example, a single emission at NLP can be either soft, or collinear, and we have
\begin{align} \label{eq:classes1}
\text{soft:} \qquad &\kappa=s\,, \qquad m(\kappa) =2\,, \\
\text{collinear:} \qquad &\kappa=c\,, \qquad m(\kappa)=1 \,.\nn
\end{align}
For a more detailed discussion see \cite{Moult:2016fqy,Moult:2017jsg}. By demanding cancellation of poles in $1/\epsilon$, as is required for an infrared and collinear safe observable, one can derive relations between contributions involving different numbers of hard, collinear and soft particles, which were used in \cite{Moult:2016fqy,Moult:2017jsg} to simplify the NNLO fixed order calculation of the NLP leading logarithms. In particular, in \cite{Moult:2016fqy,Moult:2017jsg}, it was shown that the complete result for the leading logarithms for thrust can be written at any order purely in terms of the $N$-loop hard-collinear coefficient describing a single collinear splitting
\begin{align}\label{eq:constraints_final}
\frac{1}{\sigma_0}\frac{\df\sigma^{(2,N)}}{\df\tau}
&= c_{hc,2N-1} \log^{2N-1} \tau +\cdots
\,.\end{align}
Here the dots denote subleading logarithms.
More precisely, here $c_{hc,2N-1}$ is the result for the leading $1/\epsilon$ divergence (as in \Eq{eq:constraint_setup}) with $N-1$ hard loops correcting a single collinear splitting. One class of diagram that contributes is 
\begin{align}
\fd{5cm}{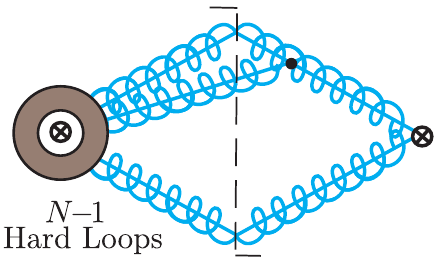}\,,\nn
\end{align}
but there will also be hard loop corrections to the amplitudes on both sides of the cut.
This relation will allow us to check our result obtained from renormalization group evolution to $\cO(\alpha_s^3)$ by expanding known results for $H\to ggg$ at two loops \cite{Gehrmann:2011aa}. In addition, it will also allow us to use our result for the all orders logarithms in thrust derived from RG evolution to understand the subleading power collinear limit at higher orders.

\subsection{General Structure}\label{sec:split_gen_structure}

Before presenting our result for the expanded amplitude squared in the collinear limit, we begin by reviewing the known IR structure of amplitudes, which we will use to organize our result. The IR structure of amplitudes is summarized by the dipole formula~\cite{Catani:1998bh} and its generalization~\cite{Dixon:2008gr,Becher:2009qa,Gardi:2009qi,Almelid:2015jia}, which provides a prediction for all the IR $1/\epsilon$ poles of scattering amplitudes at $n$ loops~(recall that we use $\alpha_s/(4 \pi)$ as the loop expansion parameter). Here we only need the full QCD amplitude for $H\to$ three partons at $n$-loops
\begin{align}
M^{(n)}=M^{(n)}_{\text{dipole}} + M^{(n)}_{R}\,.
\end{align}
Here $M^{(n)}_{\text{dipole}} $ contains all $1/\epsilon$ poles, while the remainder part $M^{(n)}_{R}$ is finite but still carries functional dependence on the kinematics that can become singular in certain limits (it is typically called the finite term but we will not use this naming scheme here). When integrating over these regions of phase space, $M^{(n)}_{R}$ must be known to all orders in $\epsilon$, and does contribute to the LL result.  More explicitly, at one-loop, we have
\begin{align}
M^{(1)} = I^{(1)}(\epsilon) M^{(0)} + M^{(1)}_{R} \,.
\end{align}
Here $I^{(1)}(\epsilon)$ is an operator in color space that can be predicted from the infrared structure of the scattering process. Using the color-charge operator notation, $I^{(1)}(\epsilon)$ can be written as~\cite{Catani:1998bh}
\begin{align}
  \label{eq:Ione}
  I^{(1)}(\epsilon) = \frac{\alpha_s}{4\pi} \frac{e^{-\epsilon \gamma_E}}{\Gamma(1 - \epsilon)} \sum_{i} \frac{1}{\mathbf{T}_i^2} \left(\mathbf{T}_i^2 \frac{1}{\epsilon^2} + \gamma_i \frac{1}{\epsilon} \right) \sum_{j\neq i} \mathbf{T}_i \cdot \mathbf{T}_j \left(\frac{\mu^2 e^{-i \pi}}{2 p_i \cdot p_j} \right)^\epsilon \,,
\end{align}
where $\mathbf{T}_i$ is the color-charge operator of massless parton $i$, $\gamma_i$ is the associated quark/gluon anomalous dimension, and we assume all QCD partons are outgoing for simplicity. 
In this paper, we have focused only on deriving a leading logarithmic result for thrust at subleading power. One obvious source of leading logarithmic contributions comes from the leading divergent terms in the amplitudes~\cite{Moult:2016fqy,Moult:2017jsg}, which exponentiate trivially. For $H \to g(p_1) g(p_2) g(p_3)$ in pure glue QCD, we have
\begin{align}\label{eq:Catani_exp}
 M_\text{dipole,LL} = \exp\left[ - \frac{\alpha_s}{4 \pi} \frac{C_A}{\epsilon^2} \left(\left(-\frac{\mu^2}{s_{12}}\right)^\epsilon 
+ \left(-\frac{\mu^2}{s_{13}}\right)^\epsilon + \left(-\frac{\mu^2}{s_{23}}\right)^\epsilon \right)  \right] M^{(0)}\,,
\end{align}
where $s_{ij} = (p_i + p_j)^2$. The subscript LL denotes that only terms contributing to thrust at LL are kept. Note that Eq.~\eqref{eq:Catani_exp} contains not only divergent terms, but also finite terms through the expansion of $\epsilon$.
After squaring the amplitudes and integrating over the phase space, the leading divergences at ${\cal O}(\alpha_s^{n+1})$ become $\alpha_s^{n+1}/\epsilon^{2 n + 1}$ at NLP, and give rise to leading logarithms for the thrust cross section. In general, the remainder part $M_R$ are not known to exhibit an iterative structure to all orders.

Typically, LL resummation at LP is carried out either by using the coherent branching formalism~\cite{Webber:1983if,Marchesini:1983bm,Marchesini:1987cf} which makes use of strongly ordered real radiation, or by computing anomalous dimensions from virtual ultraviolet divergences to hard, jet, and soft functions in SCET. However, by consistency this LL resummation also provides interesting information about higher order virtual loop corrections to a single collinear splitting.  In the next section we discuss this at both LP and NLP.  Further details for the leading power case can be found in \App{sec:LLfromCollinear}.  For this analysis both the dipole and remainder terms contribute.  Although the remainder terms do not have explicit poles in $\epsilon$, they do not necessarily vanish in the soft or colllinear limits, and in particular contain logarithms in these limits.  We will use our all orders understanding of the leading logarithms for thrust derived in Sec.~\ref{sec:fact} to show that the remainder terms also exhibit interesting exponentiation patterns.

\subsection{Subleading Power Collinear Splitting}\label{sec:split_oneloop}

To perform the expansion of the squared amplitudes in the collinear limits, we use the results of \cite{Gehrmann:2011aa}. These are in a particularly convenient form for our purposes, namely they are already expressed in a decomposition into the dipole and remainder terms.

For $H \to g(p_1)g(p_2)g(p_3)$, the collinear power expansion at amplitude level is controlled by $s = P^2 = (p_1 + p_2)^2$, the invariant mass of a pair of gluons. 
At tree level, the leading power result is given by
\begin{align}
  \label{eq:33}
  |M^{(0,0)}|^2 = 2 \tilde{\lambda}^2 \frac{(1-z+z^2)^2}{z
  (1-z)} \frac{Q^2}{s} \,,
\end{align}
where $\tilde{\lambda}^2 = 128 N_c \lambda^2 \pi^2$, $\lambda$ is the effective coupling of dimension $5$ Higgs-gluon-gluon operator, and $z$ is the longitudinal momentum fraction of $p_1$ with respective to $P$ in the collinear limit.
The next-to-leading power collinear expansion is
\begin{align}
  \label{eq:34}
  |M^{(2,0)}|^2 = 2 \tilde{\lambda}^2 \frac{1 + 2 z - 3 z^2 + 2 z^3 - z^4}{z (1-z)} \,.
\end{align}
Here we have used a double superscript notation where the first superscript indicates the power in $s/Q^2$, and the second indicates the order in $\alpha_s$. Eq.~\eqref{eq:34} contains end-point singularity in the momentum fraction, which is regularized by the $d$ dimension phase space measure. 
For the purpose of extracting the leading logarithms, it is only
necessary to consider the $z\to 0$ or $z\to 1$ limit. In the current
case the two limits are identical, and we find
\begin{align}
  |M^{(2,0)}|_\text{LL}^2   =  \tilde{\lambda}^2 \frac{2}{z
   (1-z)}  \,,
\end{align}
where we use subscript LL to denote that only the end-point singular term in $z$ is retained.
We can use these to define the tree level LP and NLP splitting functions, valid at LL level
\begin{align}\label{eq:split_def}
P_{gg,\text{LL}}^{(0,0)}\ & = \frac{Q^2}{s} \frac{2}{z(1-z)}\,, \quad
P_{gg,\text{LL}}^{(2,0)}\ = \frac{2}{z(1-z)}\,.
\end{align}
Here we see the explicit suppression in $s/Q^2$ of the NLP result.
We then have
\begin{align}
  \label{eq:35}
  |M^{(0,0)}|_\text{LL}^2 & = \tilde{\lambda}^2  P_{gg,\text{LL}}^{(0,0)}  \,, \qquad
  |M^{(2,0)}|_\text{LL}^2 =  \tilde{\lambda}^2  P_{gg,\text{LL}}^{(2,0)}\,.
\end{align}

Using \Eq{eq:Catani_exp} it is trivial to give the all loop result for squared amplitude for the terms predicted by dipole formula. We find
\begin{align}
|M|^2_\text{dipole,LP,LL}&\!\! = \tilde{\lambda}^2  P_{gg,\text{LL}}^{(0,0)}\exp\left( {F_{\text{dipole}}} \right)\,, \qquad
|M|^2_\text{dipole,NLP,LL} \!\!= \tilde{\lambda}^2  P_{gg,\text{LL}}^{(2,0)}\exp\left( {F_{\text{dipole}}} \right)\,,
\end{align}
where
\begin{align}
\label{eq:dipole}
F_{\text{dipole}}= \frac{\alpha_s \mu^{2 \e}}{4 \pi}  \frac{(-2
  C_A)}{\e^2} \left( [(1-z)Q^2]^{-\e} + s^{-\e} + [z Q^2]^{-\e}\right)\,.
\end{align}
Interestingly, the form of the dipole term guarantees that its leading logarithmic loop corrections are independent of the power expansion. The power expansion arises only in the expansion of the tree level amplitude squared.

Much more interesting are the remainder terms of the amplitude, whose all order form is not predicted. We can begin by looking at their form at one-loop.
By inspecting the higher order in $\e$ terms in the
remainder term of the amplitude, we can write down an all-order-in-$\e$ expression
for the leading transcendental piece of the remainder terms (i.e. the piece required to give the LL for thrust). We find
\begin{align}
  \label{eq:47}
       2 \mathrm{Re}\big[
&\,  M^{(0)*}M_R^{(1)}\big]\Big|_\text{LP,LL} = 
 - 2 C_A \tilde  \lambda^2  P_{gg,\text{LL}}^{(0,0)}  
\nn\\ &\,
\times \frac{\alpha_s\,\mu^{2 \epsilon} }{4\pi}  \Bigg[\left(\frac{[Q^2]^{-\e}}{\e^2} - \frac{[z (1-z) Q^2]^{-\e}}{\e^2}
  \right) - \left(\frac{[s]^{-\e}}{\e^2} - \frac{[z(1-z)
  s]^{-\e}}{\e^2}  \right)\Bigg]\,.
\end{align}
The structure of this leading transcendental component of the remainder term is quite interesting. Expanding it, we see that both the $1/\epsilon^2$ and $1/\epsilon$ poles cancel, giving a finite result
\begin{align}
&\Bigg[\left(\frac{[Q^2]^{-\e}}{\e^2} - \frac{[z (1-z) Q^2]^{-\e}}{\e^2}
  \right) - \left(\frac{[s]^{-\e}}{\e^2} - \frac{[z(1-z)
  s]^{-\e}}{\e^2}  \right)\Bigg] \nn \\
&= \frac{[Q^2]^{-\epsilon}  }{2} \left(  -\log^2 \left( \frac{s}{Q^2} \right)  -\log^2(z(1-z)) +\log^2 \left( \frac{s(1-z)z}{Q^2}   \right) +\cO(\epsilon)  \right)\,.
\end{align}
However, we see that this term secretly contains leading poles in $1/\epsilon$ when written in the form of \Eq{eq:constraint_setup} and therefore will contribute to the LL result at LP. The reason is that when integrating over the momentum fraction $z$ using $d$ dimension phase space measure, there is a mismatch in the exponent of $z$ between different terms. Since this is a non-traditional way to obtain the leading logarithms for the thrust distribution, we provide a more detailed explanation in Appendix~\ref{sec:LLfromCollinear}. For the NLP terms, we find the exact same structure, with only a different prefactor
\begin{align}
  \label{eq:48}
         2 \mathrm{Re}\big[
&\,  M^{(0)*}M_R^{(1)}\big]\Big|_\text{NLP,LL} = - 2 C_A
\tilde   \lambda^2  P_{gg,\text{LL}}^{(2,0)}  
\nn\\ &\,
\times \frac{\alpha_s\,\mu^{2 \epsilon} }{4\pi}  \Bigg[\left(\frac{[Q^2]^{-\e}}{\e^2} - \frac{[z (1-z) Q^2]^{-\e}}{\e^2}
  \right) - \left(\frac{[s]^{-\e}}{\e^2} - \frac{[z(1-z)
  s]^{-\e}}{\e^2}  \right)\Bigg]\,.
\end{align}
Interestingly, as was the case for the dipole terms, we again see that the transcendental structure is the same at LP and NLP, and it just multiplies the tree level splitting function.

Going to two loops, quite interestingly, we find that the remainder term is
\begin{align}
  \label{eq:49}
& \,  2 \mathrm{Re}\big[ M^{(0)*}M_R^{(2)}\big] +
  M_R^{(1)*}M_R^{(1)}\Big|_\text{LP,LL} =  
 2 C_A^2 \tilde  \lambda^2  P_{gg,\text{LL}}^{(0,0)}   
\nn\\ &\qquad \times \frac{\alpha_s^2\,\mu^{4 \epsilon} }{(4\pi)^2} 
 \Bigg[\left(\frac{[Q^2]^{-\e}}{\e^2} - \frac{[z (1-z) Q^2]^{-\e}}{\e^2}
  \right) - \left(\frac{[s]^{-\e}}{\e^2} - \frac{[z(1-z)
  s]^{-\e}}{\e^2}  \right)\Bigg]^2\,.
\end{align}
Note that only the $\Ord{\e^0}$ terms in Eq.~\eqref{eq:49} are
explicitly verified using the two-loop amplitudes. To verify to higher
order in $\e$, one needs to know the two-loop amplitudes also to higher order in $\e$, which are currently not available in the
literature. However, since this contribution is related to the hard-collinear contribution in the effective theory, the renormalizability of the effective theory guarantees that the all loop result can be obtained by RG evolution of the lowest order result. To LL, the power expansion in the amplitudes acts only on the kinematic factors giving rise to the lowest order splitting functions in \Eq{eq:split_def}, but not on the transcendental function. This therefore fixes the all order in $\e$ form of Eq.~\eqref{eq:49}. Compared with Eq.~\eqref{eq:47}, we have the relation
\begin{align}
  \label{eq:50}
  & \, \frac{2 \mathrm{Re}\big[ M^{(0)*}M_R^{(2)}\big] +
  M_R^{(1)*}M_R^{(1)} \Big|_\text{LP,LL}}{|M^{(0,0)}|_\text{LL}^2} =  
\frac{1}{2!} \left( \frac{2 \mathrm{Re}\big[
    M^{(0)*}M_R^{(1)}\big]\Big|_\text{LP,LL}}{|M^{(0,0)}|_\text{LL}^2}\right)^2\,,
\end{align}
that is, the remainder term also exponentiates. Similarly, for the
NLP piece, we have
\begin{align}
  \label{eq:51}
    & \, \frac{2 \mathrm{Re}\big[ M^{(0)*}M_R^{(2)}\big] +
  M_R^{(1)*}M_R^{(1)} \Big|_\text{NLP,LL}}{|M^{(2,0)}|_\text{LL}^2} =  
\frac{1}{2!} \left( \frac{2 \mathrm{Re}\big[
      M^{(0)*}M_R^{(1)}\big]\Big|_\text{NLP,LL}}{|M^{(2,0)}|_\text{LL}^2} \right)^2\,.
\end{align}
Here we observe exponentiation of the remainder term at LP and NLP, and furthermore, we again see that the transcendental structure at both LP and NLP is identical. 

With the expanded result for the squared amplitude, we can simply integrate it over the collinear phase space to obtain the result for thrust. We find
\begin{align}
  \label{eq:52}
  \frac{1}{\sigma_0} \frac{\df\sigma^{(2)}}{\df \tau} =
 \left( \frac{\alpha_s}{4\pi} \right) 8 C_A \log\tau -
  \left(\frac{\alpha_s}{4\pi}\right)^2
32 C_A^2 \log^3 \tau + \left(\frac{\alpha_s}{4\pi}\right)^3 64 C_A^3
  \log^5 \tau +\cO(\alpha_s^4) \,.
\end{align}
This agrees with the result derived from the RG in \Eq{eq:resum_nlp}, and provides an explicit check at $\cO(\alpha_s^3)$ of the result from the RG. The terms  to $\cO(\alpha_s^2)$ were also computed in \cite{Moult:2017jsg} using this technique. The $\cO(\alpha_s^3)$ term has not previously appeared in the literature.

We can now use the higher order terms predicted by the RG to study the collinear limit at higher loop orders. In particular, since we have derived using the RG that the leading logarithms for thrust exponentiate into a Sudakov, given in \Eq{eq:resum_nlp}, the all-loop expansion of the amplitudes in the collinear limit must agree with this exponentiation.

We have already shown that at least to two loops, the leading logarithmic contribution of the remainder terms exponentiate.
Combined with the exponentiation of the dipole terms, we conjecture that to all orders, amplitudes in the collinear limit through to NLP exponentiate, namely
\begin{align}\label{eq:split_allorders}
\left. \left[ M^* M \right]\right |_{\text{LP,LL}}&=  \tilde \lambda^2 P_{gg,\text{LL}}^{(0,0)} e^{F_\text{dipole} + F_R}  \,,
\quad
\left. \left[ M^* M \right]\right |_{\text{NLP,LL}}=  \tilde \lambda^2 P_{gg,\text{LL}}^{(2,0)} e^{F_\text{dipole} + F_R}  \,,
\end{align}
where 
\begin{align}
\label{eq:remainder}
F_R=\frac{\alpha_s \mu^{2 \epsilon}}{4 \pi}  (-2 C_A)  \left[ \left(\frac{[Q^2]^{-\e}}{\e^2} - \frac{[z (1-z) Q^2]^{-\e}}{\e^2}
  \right) - \left(\frac{[s]^{-\e}}{\e^2} - \frac{[z(1-z)
  s]^{-\e}}{\e^2}  \right) \right]\,.
\end{align}
In particular, this result reproduces the leading logarithms in thrust obtained through RG evolution to all loop order in \Eq{eq:resum_nlp}. Note that this is an amplitude level statement, and while we have explicitly checked it to two loops, and when integrated over $z$ it agrees with our result obtained from the RG for thrust, which provides a strong check, we phrase it only as a conjecture, since it is possible $z$ dependent terms that do not give rise to leading logarithms for the thrust observable could be present.
This seems to imply an interesting iterative structure for the remainder terms of the amplitude, which is relevant for leading logarithmic resummation, and goes beyond the dipole formula.  This would be interesting to investigate further, and we hope that the study of subleading power limits will lead to a further understanding.

Here we have only considered the case of $H\to ggg$, but it is important to understand the universality of the above subleading power splitting functions, and in particular of their loop corrections, even at a given logarithmic accuracy. The universality of subleading power collinear factorization has been studied at tree level in \cite{Nandan:2016ohb}, but it would be interesting to try to extend it to all loop order using the techniques in this paper. A perhaps related question is the definition of an infrared finite remainder function in planar ${\cal N}=4$ SYM, where a clever definition of exponentiated terms can lead to a better behaved remainder function~\cite{Caron-Huot:2016owq}. 

\section{Conclusions}\label{sec:conc}

In this paper we have, for the first time, resummed to all orders in $\alpha_s$  subleading power logarithms for the thrust observable to LL accuracy for pure glue $H\to gg$. We have shown that the subleading power logarithms exponentiate to all orders into a Sudakov exponential controlled by the cusp anomalous dimension multiplying a  logarithm, see \Eq{eq:resum_nlp}. Resummation is achieved by RG evolution of gauge invariant non-local Wilson line operators and its accuracy is systematically improvable. 

The renormalization of subleading power jet and soft functions requires the introduction of a new class of universal soft and collinear functions, which we termed $\theta$-jet and $\theta$-soft functions. These functions, which involve $\theta$-functions of the measurement, appear through operator mixing, and we argued that they will play a general role in renormalization and resummation at subleading powers. We introduced a simple example which allowed us to understand the structure of these functions to all orders in $\alpha_s$, as well as to derive their renormalization group evolution, which we proved closes into a $2\times2$ mixing equation. We analytically solved this subleading power RG mixing equation, including the effects of running coupling.

We checked our result derived from RG evolution to $\cO(\alpha_s^3)$ by direct calculation of the power corrections. Using consistency relations from the cancellation of IR poles, the leading logarithms can be derived entirely from the collinear limit, allowing us to use our all orders result derived from the RG equations to understand higher order loop corrections to the subleading power collinear limit. We showed explicitly that to two-loops all leading transcendental pieces in the collinear and subleading power collinear limit exponentiate. We conjectured that this exponentiation holds to all loop order, and showed that this results in agreement with the results for the thrust observable derived from RG evolution. This seems to indicate an interesting structure for the IR finite terms in the subleading power collinear limits, beyond what is predicted by the dipole formula, and it would be interesting to investigate this further.  

Since this represents the first all orders resummation of NLP logarithms for an event shape, there are many interesting directions in which it can be extended. In particular,  it will be important to extend our results to higher logarithmic accuracy to understand what universal structures persist. The simplicity of the leading logarithmic structure to all powers suggests the possibility of a simple structure. It will also be interesting to study subleading power corrections for other observables, such as $q_T$ or in the threshold limit, as well as to extend the calculation to the $N$-jet case, for example for the $N$-jettiness observable \cite{Stewart:2010tn}. The renormalization of amplitude level hard scattering operators for the $N$-jet case was recently considered \cite{Beneke:2017ztn}, which provides an important ingredient in this direction. Our work provides a path for the systematic resummation of subleading power logarithms for event shapes, and we hope that this will lead to an improved understanding of the all orders structure of the subleading power soft and collinear limits.

\begin{acknowledgments}
This work was supported in part by the Office of Nuclear Physics of the U.S.
Department of Energy under Contract No. DE-SC0011090, by the Office of High Energy Physics of the U.S. Department of Energy under Contract No. DE-AC02-05CH11231, by the Simons Foundation Investigator Grant No. 327942, and by the LDRD Program of LBNL. IM thanks Zhejiang University and the MIT Center for Theoretical Physics for hospitality while portions of this work were performed.
\end{acknowledgments}

\appendix

\section{Solution to Subleading Power RG Mixing Equation in Momentum Space}\label{sec:inversefourier}

In \Sec{sec:solution} we have shown that in the leading log approximation, and in the case when $\Gamma^{(0)}_{11}=\Gamma^{(0)}_{22}$, the solution to the subleading power RG mixing equation in position space is \Eq{eq:FLL}. Here we provide additional details on the transformation of this result back to momentum space. In position space the logarithms for the boundary condition are minimized by the choice $\mu_0=\mu_y$. For thrust at subleading power there are no distributions, and the logarithms have a simple correspondence between position and momentum space without subtleties. This is analogous to the situation between position space and cumulative thrust at leading power. To derive an exact relation for the Fourier transform we note that 
\begin{align} \label{eq:FTNLP}
 \int\! \frac{dy}{2\pi}\, e^{i k y}\, (iy)^{-1-\epsilon} = \frac{ \theta(k) k^\epsilon}{\Gamma(1+\epsilon)} \,,
\end{align}
where branch cuts are defined by $y=y-i0$.
Defining $e^{-\epsilon\gamma_E}/\Gamma(1+\epsilon) = \sum_{k=0}^\infty e_k\, \epsilon^k$, we have $e_{0}=1$, $e_{1}=0$, $e_{2}=-\zeta_2/2$, $e_3=\zeta_3/2$, etc. Expanding \Eq{eq:FTNLP} in $\epsilon$ leads to the identity we need to connect the subleading power logarithms in position and momentum space,
\be\label{eq:fourier}
	\int \frac{dy}{2\pi} e^{iky}\frac{\log^n(iye^{\gamma_E} \mu^p)}{i(y-i0)} = (-1)^n \sum_{j = 0}^{n} \frac{n!}{j!}\, e_{n-j}\, \log^j \Bigl(\frac{k}{\mu^p}\Bigr)\, \theta(k)\,.
\ee
Keeping only the LL term on the RHS gives the simple correspondence $\log^n(iye^{\gamma_E}\mu^p)/(iy) \to (-1)^n \log^n(k/\mu^p)\, \theta(k)$. 
To see how this works in an explicit example, we can rewrite the resummed position space result in \Eq{eq:FLL} as
\begin{align}\label{eq:Fposspace}
	\tilde F^{(2)\LL}_{\delta}(y,\mu)
 &= \tilde U_{\delta\theta}^{F,{\rm LL}}(y,\mu,\mu_0) \tilde F_{\theta}^{(2)}(y,\mu_0)
  = A \frac{\left(e^{\gamma_E}iy\mu_0^p \right)^{\omega}}{i(y-i0)} \\
 &= \frac{A}{i(y-i0)} e^{\omega \log\left(e^{\gamma_E}iy \mu_0^p\right)}
   = A \sum_{n=0}^\infty \frac{1}{n!} \omega^n \frac{\log^n\left(e^{\gamma_E}iy\mu_0^p\right)}{i(y-i0)} \,,\nn
\end{align}
where $A \equiv -\frac{\gamma^{(0)}_{12}}{2\beta_0}\log r\exp\Bigl[\frac{p\pi\Gamma^{(0)}_{11}}{\beta_0^2\alpha_s(\mu_0)} \left(\frac{1}{r} - 1 + \log r \right) \Bigr]$ and  $\omega\equiv -\frac{\Gamma^{(0)}_{11}}{2\beta_0}\log(r) $ are dimensionless $y$ independent expressions, where here $r=\alpha_s(\mu)/\alpha_s(\mu_0)$. Using \Eq{eq:fourier} we have
\begin{align}
	F^{(2)\LL}_{\delta}(k,\mu) &= \int \frac{dy}{2\pi} e^{iky} \tilde F^{(2)\LL}_{\delta}(y,\mu)  = A \sum_{n=0}^{\infty} \frac{1}{n!} \omega^n  \int \frac{dy}{2\pi} e^{iky} \frac{\log^n\left(e^{\gamma_E}iy\mu_0^p \right)}{i(y-i0)}  
  \nn \\
	&= A \sum_{n=0}^{\infty} \sum_{j= 0}^{n} \omega^n (-1)^n\frac{e_{n-j}}{j!} \log^j\Bigl(\frac{k}{\mu_0^p}\Bigr) \theta(k)  
  \,.
\end{align}
Here all the terms with $j < n$ are subleading logs, therefore at LL order we can keep just the $j = n$ term to give
\begin{align}\label{eq:Fmomspace}
	F^{(2)\LL}_{\delta}(k,\mu) &= A \sum_{n=0}^{\infty} \frac{(-\omega)^n}{n!} e_{0} \log^n(k) \theta(k) = A e^{-\omega\log(k)} \theta(k)   \\
	&= -\theta(k) \frac{\gamma^{(0)}_{12}}{2\beta_0}\log r\exp\left[\frac{p\pi\Gamma^{(0)}_{11}}{\beta_0^2\alpha_s(\mu_0)} \left(\frac{1}{r} - 1 + \log r \right) \right] \left(\frac{k}{\mu_0^p}\right)^{\frac{\Gamma^{(0)}_{11}}{2\beta_0}\log(r) } \nn\\
	&\equiv \theta(k) U_{\delta \theta}^\LL(k,\mu,\mu_0) \,.\nn
\end{align}
Note that this is simply obtained from the starting result in \Eq{eq:Fposspace} by taking $iye^{\gamma_E}\to 1/k$ everywhere, except for in the explicit prefactor $1/(y-i0)\to \theta(k)$. 
\Eq{eq:Fmomspace} is the LL solution to the subleading RG mixing equation in momentum space which was quoted in the main text in \Eq{eq:Umomspace}.

\section{Leading Logarithms for Thrust from Collinear Limits of Amplitudes}\label{sec:LLfromCollinear}

In this Appendix we explain how to obtain the LP LL series for thrust using only the information from collinear limits of scattering amplitudes. The NLP case, which is the focus of this paper, is similar. However, here we present the LP case in detail since this approach to obtaining the LL series is not traditional. The key idea is that the infrared scale dependence should cancel out in a physical cross section. Just as in the NLP analysis leading to \Eq{eq:constraints_final}, consistency at LP implies that the LL term can be obtained from loop corrections to the amplitude for a single collinear emission encoded in coefficients $d_{hc,2N}^{(0)}$, 
\begin{align}
\frac{1}{\sigma_0}\frac{\df\sigma^{(0,N)}}{\df\tau}
&= d_{hc,2N}^{(0)} \frac{\log^{2N - 1} \tau}{\tau} +\cdots
\,.\end{align}
We will work this out explicitly for the first two loop orders below.

Here, as in the text, we take thrust for Higgs decay in pure glue QCD as an example. We write the NLO cumulant at LP as
\begin{align}
  \label{eq:cumulantNLO}
  R^{(0,1)}(\tau) & = \frac{1}{\sigma_0}\int_0^\tau d\tau' \frac{d\sigma^{(0,1)}}{d\tau'} 
\nn\\
& =\frac{\alpha_s}{4 \pi} \frac{C_A}{\e^2} \left(c_h \left( \frac{\mu^2}{Q^2} \right)^\e + c_c \left( \frac{\mu^2}{\tau Q^2} \right)^\e  + c_s \left( \frac{\mu^2}{\tau^2 Q^2} \right)^\e\right) + {\cal O}\left(\frac{1}{\e} \right) \,, 
\end{align}
where we have separated the contribution between hard virtual corrections $c_h$, collinear corrections $c_c$, and soft corrections $c_s$. For a physical cross section both the divergent terms and the LL $\mu$ dependence should cancel. In particular, they should cancel between the $1/\e^2$ terms in Eq.~\eqref{eq:cumulantNLO}. There is no cancellation between the expansion of the $1/\e^2$ terms and the ${\cal O}(1/\e)$ terms. That's why we don't need to write down the ${\cal O}(1/\e)$ terms explicitly, at least for LL. It then follows that
\begin{align}
  \label{eq:NLOconsistency}
  c_h = - \frac{1}{2} c_c \,, \qquad c_s  = - \frac{1}{2} c_c \,.
\end{align}
Substituting the relation in Eq.~\eqref{eq:NLOconsistency} into Eq.~\eqref{eq:cumulantNLO}, we find
\begin{align}
  \label{eq:NLOLL}
  R^{(0,1)}(\tau) = - \frac{1}{2} \frac{\alpha_s}{4 \pi} \, C_A c_c \log^2 \tau  + \text{subleading logs}\,.
\end{align}
That is, the leading logarithm at NLO is uniquely determined by the contribution from the hard collinear splitting. Specifically, at NLO for thrust, the collinear corrections to the cumulant can be written as
\begin{align}
  \label{eq:RcNLO}
  R_{c,\text{LL}}^{(0,1)}(\tau) & = 2 \frac{\alpha_s \mu^{2 \e}}{4 \pi} \int_0^{\tau Q^2} \frac{ds}{Q^2} \int_0^1 dz \frac{e^{\e \gamma_E} [s z (1-z) ]^{-\e}}{\Gamma(1-\e)} C_A P_{gg,\text{LL}}^{(0,0)}
\nn\\
& = \frac{\alpha_s}{4 \pi}  \frac{8 C_A}{\e^2} \left(\frac{\mu^2}{\tau Q^2}\right)^\e+ {\cal O}\left( \frac{1}{\e} \right) \,,
\end{align}
where $P_{gg,\text{LL}}^{(0,0)}$ is introduced in Eq.~\eqref{eq:split_def}.
Therefore $c_c = 8$, and $R^{(0,1)}(\tau) = - \frac{\alpha_s}{\pi} C_A \log^2\tau + \text{subleading logs}$. 

At NNLO, there are several combinations of different modes, but the idea is similar. We write the cumulant as
\begin{align}
  \label{eq:RNNLO}
  R^{(0,2)} (\tau) & =\left( \frac{\alpha_s}{4 \pi}  \right)^2 \frac{C_A^2}{\e^4}
\left(c_{hh} \left( \frac{\mu^4}{Q^4} \right)^{\e} + c_{hc} \left( \frac{\mu^4}{\tau Q^4} \right)^\e  + (c_{cc} + c_{hs}) \left( \frac{\mu^4}{\tau^2 Q^4} \right)^\e + c_{cs} \left( \frac{\mu^4}{\tau^3 Q^4}  \right)^\e
\right. \nn\\
&
\left. + c_{ss} \left( \frac{\mu^4}{\tau^4 Q^4} \right)^\e \right) + {\cal O}\left(\frac{1}{\e^3} \right) \,, 
\end{align}
Here $c_{hh}$ denotes hard modes contributions from pure virtual diagrams, $c_{hc}$ denotes real-virtual contributions with virtual hard mode and real collinear mode, $c_{cc}$ denotes  both real-virtual or double real contributions with virtual or real collinear modes, $c_{hs}$ denotes real-virtual contributions with virtual hard mode and real soft mode, and finally $c_{ss}$ denotes real-virtual or double real contributions with virtual or real soft modes. 
Demanding that all the poles and $\mu$ dependence from expanding the $1/\e^4$ terms cancel, we find
\begin{gather}
  \label{eq:NNLOconsistency}
  c_{hc} = -4 c_{hh} \,, \qquad c_{cc} + c_{hs} = 6 c_{hh} \,, \qquad c_{cs} =  - 4 c_{hh} \,, \qquad  c_{ss} = c_{hh} \,.
\end{gather}
We then find
\begin{align}
  \label{eq:RNNLOhc}
  R^{(0,2)}(\tau) = - \left( \frac{\alpha_s}{4 \pi}  \right)^2 \frac{C_A^2}{4} c_{hc} \log^4 \tau + \text{subleading logs} \,.
\end{align}
Specifically, the real-virtual collinear corrections to the cumulant is given by
\begin{align}
  \label{eq:RVcollinear}
  R_{RVc,\text{LL}}^{(0,2)}(\tau) & = 2 \frac{\alpha_s \mu^{2 \e}}{4 \pi} \int_0^{\tau Q^2} \frac{ds}{Q^2} \int_0^1 dz \frac{e^{\e \gamma_E} [s z (1-z) ]^{-\e}}{\Gamma(1-\e)} C_A P_{gg,\text{LL}}^{(0,0)} \left( F_\text{dipole} + F_R \right) \,,
\end{align}
where we have separated the corrections into the dipole term and the remainder term, see Eq.~\eqref{eq:dipole} and \eqref{eq:remainder}. The dipole term gives
\begin{align}
  \label{eq:Rdipole}
  R_{RVc,\text{dipole,LL}}^{(0,2)}(\tau) & = \left(\frac{\alpha_s}{4 \pi} \right)^2 \left[ - \frac{24 C_A^2}{\e^4} \left(\frac{\mu^4}{\tau Q^4} \right)^\e -  \frac{8 C_A^2}{\e^4} \left(\frac{\mu^4}{\tau^2 Q^4} \right)^\e \right]+ {\cal O}\left( \frac{1}{\e^3} \right)\,.
\end{align}
And the remainder term gives
\begin{align}
  \label{eq:Rremainder}
  R_{RVc,R,\text{LL}}^{(0,2)}(\tau) & = \left(\frac{\alpha_s}{4 \pi} \right)^2 \left[- \frac{8 C_A^2}{\e^4}  \left(\frac{\mu^4}{\tau Q^4} \right)^\e + \frac{4 C_A^2}{\e^4}  \left(\frac{\mu^4}{\tau^2 Q^4} \right)^\e \right] + {\cal O}\left( \frac{1}{\e^3} \right)\,.
\end{align}
Adding the dipole and remainder terms, we find that the hard-collinear coefficient is $c_{hc} = -32$, and the NNLO cumulant is
\begin{align}
  \label{eq:RNNLOres}
  R^{(0,2)}(\tau) = \left( \frac{\alpha_s}{4 \pi} \right)^2 8 \, C_A^2 \log^4 \tau + \text{subleading logs} \,.
\end{align}
This is the correct leading logarithm for thrust. We see explicitly that both the dipole term and the remainder term contribute to thrust at LL. The analysis above can be straightforwardly carried out to all orders in $\alpha_s$.

\bibliography{subRGE}
\bibliographystyle{JHEP}

\end{document}